\documentclass{llncs}
 
\usepackage{array}
\usepackage[dvips]{graphicx}
\usepackage{epic, eepic}
\usepackage{amssymb,amsmath}
\usepackage{latexsym}
\usepackage{subfigure}
\usepackage[usenames]{color}
\usepackage{multirow}
\usepackage{gastex}
\usepackage{subfigure}
\newcounter{compressEnum}
\renewcommand{\thecompressEnum}{$\roman{compressEnum}$}
\newenvironment{compressEnum}
{\setcounter{compressEnum}{0}}
{}
\newcommand{\itCompress}{\stepcounter{compressEnum}{(\thecompressEnum) }}

\newenvironment{myProof}[1][\unskip]{\medskip\par\noindent{\bfseries Proof
    #1.}\ \ 
        \global\def\qed{\origQED\global\def\qed{}}\penalty10000}%
        {\qed\par\medskip\global\def\qed{\origQED\global\def\qed{}}}

\def\endof{%
  \leavevmode
  \parfillskip=0pt%
  \widowpenalty=10000%
  \displaywidowpenalty=10000%
  \finalhyphendemerits=0%
  \unskip\nobreak\null\hfil\penalty50\hskip2em\null\hfill%
}
\def\eodsymbol{\ensuremath\square}
\def\eopsymbol{\ensuremath\blacksquare}

\def\origEOD{\nobreak\leavevmode\endof\eodsymbol\par}
\def\EOD{\origEOD\global\def\EOD{}}
\def\origQED{{\nobreak\leavevmode\endof\eopsymbol\par\medskip}}
\def\qed{\origQED\global\def\qed{}}

\def\abs#1{\ensuremath{\lvert #1\rvert}}

\newcommand{\nat}{\mathbb N} 
\newcommand{\rat}{{\mathbb Q}}

\newcommand{\posreal}{{\mathbb R}^{\geq 0}}
\newcommand{\sposreal}{{\mathbb R}^{> 0}}
\newcommand{\real}{{\mathbb R}}

\newcommand{\tuple}[1]{\langle #1 \rangle}

\newcommand{\C}{\mathcal{C}}

\newcommand{\Last}{{\sf Last}}

\newcommand{\La}{{\sc Lavg}}
\newcommand{\Li}{{\sc Linf}}
\newcommand{\Ls}{{\sc Lsup}}
\newcommand{\Di}{{\sc Disc}}

\newcommand{\nmax}{{\sc NSup}}
\newcommand{\ndi}{{\sc NDisc}}
\newcommand{\nla}{{\sc NLavg}}
\newcommand{\nli}{{\sc NLinf}}
\newcommand{\nls}{{\sc NLsup}}
\newcommand{\nbw}{{\sc NBW}}

\newcommand{\dmax}{{\sc DSup}}
\newcommand{\ddi}{{\sc DDisc}}
\newcommand{\dla}{{\sc DLavg}}
\newcommand{\dli}{{\sc DLinf}}
\newcommand{\dls}{{\sc DLsup}}
\newcommand{\dbw}{{\sc DBW}}

\newcommand{\ndli}{{\sc 
\raisebox{2.0pt}{\scalebox{0.45}{\begin{tabular}{c}D\\[-3pt]N\end{tabular}}}Linf
}}
\newcommand{\ndmax}{{\sc 
\raisebox{2.0pt}{\scalebox{0.45}{\begin{tabular}{c}D\\[-3pt]N\end{tabular}}}Sup
}}
\newcommand{\nd}{\hspace{-2pt}{\sc 
\raisebox{2.0pt}{\scalebox{0.45}{\begin{tabular}{c}D\\[-2.8pt]N\end{tabular}}}\hspace{0pt}}yW
}

\newcommand{\Val}{\mathsf{Val}}
\newcommand{\Maxf}{\mathsf{Max}}
\newcommand{\Max}{\mathsf{Sup}}
\newcommand{\LimSup}{\mathsf{LimSup}}
\newcommand{\LimInf}{\mathsf{LimInf}}
\newcommand{\LimAvg}{\mathsf{LimAvg}}
\newcommand{\Disc}{\mathsf{Disc}}
\newcommand{\Avg}{\mathsf{Avg}}
\newcommand{\Sum}{\mathsf{Sum}}

\def\set#1{\ensuremath{\{#1\}}}
\newcommand{\ok}{\raisebox{0.2em}{$\sqrt{}$}}
\newcommand{\ko}{$\times$}
\newcommand{\weight}{\gamma}

\DeclareMathOperator{\op}{op}

\makeatletter

\begingroup \catcode `|=0 \catcode `[= 1
\catcode`]=2 \catcode `\{=12 \catcode `\}=12
\catcode`\\=12 |gdef|@xcomment#1\end{comment}[|end[comment]]
|endgroup

\def\@comment{\let\do\@makeother \dospecials\catcode`\^^M=10\def\par{}}

\def\begincomment{\@comment\@xcomment}

\makeatother

\newenvironment{comment}{\begincomment}{}

\title{Expressiveness and Closure Properties for Quantitative Languages} 

\author{Krishnendu Chatterjee\inst{1} \and Laurent Doyen\inst{2} \and Thomas A. Henzinger\inst{2}}

\institute{IST, Austria \\
\and EPFL, Lausanne, Switzerland
}

\begin{document}
\pagestyle{plain}
\maketitle

\begin{abstract}
Weighted automata are nondeterministic automata with numerical weights on
transitions. They can define quantitative languages~$L$ that
assign to each word~$w$ a real number~$L(w)$.
In the case of infinite words, the value of a run is naturally computed as
the maximum, limsup, liminf, limit average, or discounted sum of the
transition weights.
We study expressiveness and closure questions about these quantitative
languages.

We first show that the set of words with value greater than a threshold
can be non-$\omega$-regular for deterministic limit-average and
discounted-sum
automata, while this set is always $\omega$-regular when the threshold
is isolated (i.e., some neighborhood around the threshold contains 
no word).
In the latter case, we prove that the $\omega$-regular language is robust
against small perturbations of the transition weights.

We next consider automata with transition weights $0$ or $1$ and show that they
are as expressive as general weighted automata in the limit-average case, but not
in the discounted-sum case.

Third, for quantitative languages $L_1$ and~$L_2$, we consider the operations
$\max(L_1,L_2)$, $\min(L_1,L_2)$, and $1-L_1$, which generalize
the boolean operations on languages, as well as the sum $L_1 + L_2$.
We establish the closure properties of all classes
of quantitative languages with respect to these four operations.

\end{abstract}

\section{Introduction}
A boolean language $L$ can be viewed as a function that assigns to each word $w$ a boolean value,
namely, $L(w) = 1$ if the word $w$ belongs to the language, and $L(w) = 0$ otherwise. 
Boolean languages model the computations of reactive programs. The verification
problem ``does the program $A$ satisfy the specification $B$?'' 
then reduces to the language-inclusion problem ``is $L_A \subseteq L_B$?'',
or equivalently, ``is $L_{A}(w)\leq L_{B}(w)$ for all words~$w$?'',
where $L_A$ represents all behaviors of the program, and $L_B$ contains 
all behaviors allowed by the specification. When boolean languages are defined
by finite automata, this elegant framework
is called the \emph{automata-theoretic approach} to model-checking~\cite{VardiW86}.

In a natural generalization of this framework, a cost function
assigns to each word a real number instead of a boolean value. 
For instance, the value of a word (or behavior) can be interpreted as the amount of some resource 
(e.g., memory consumption, or power consumption) that the program needs to produce it,
and a specification may assign a maximal amount of available resource to each 
behavior, or bound the long-run average available use of the resource. 

Weighted automata over semirings (i.e., finite automata with transition weights 
in a semiring structure) have been used to define cost functions, 
called formal power series for finite words~\cite{Wautomata,KuichS86}
and $\omega$-series for infinite words~\cite{CulikK94,DrosteK03,EsikK04}.
In~\cite{CDH08}, we study new classes of cost functions using 
operations over rational numbers that do not form a semiring. 
We call them \emph{quantitative languages}.
We set the value of a (finite or infinite) word $w$ as the maximal value of all runs over~$w$
(if the automaton is nondeterministic, then there may be many runs over~$w$), 
and the value of a run $r$ is a function of the (finite or infinite) sequence 
of weights that appear along~$r$. We consider several 
functions, such as $\Maxf$
%
%
and $\Sum$ of weights for finite runs, 
and $\Max$, $\LimSup$, $\LimInf$, 
limit average, and discounted sum of weights for infinite runs.
For example, peak power consumption can be modeled as the maximum of a sequence 
of weights representing power usage; energy use can be modeled as the sum;
average response time as the limit average \cite{CCHK+05,CAHS03}.
Quantitative languages can also be used to specify and verify reliability 
requirements: if a special symbol $\bot$ is used to denote failure and has 
weight~$1$, while the other symbols have weight $0$, one can use a limit-average 
automaton to specify a bound on the rate of failure in the long run~\cite{CGHIKPS08}.
The discounted sum can be used to specify that failures happening
later are less important than those happening soon~\cite{AHM03}.

The \emph{quantitative language-inclusion problem} ``Given two automata~$A$ and~$B$, 
is $L_{A}(w)\leq L_{B}(w)$ for all words~$w$?'' can then be used to check, say, if for each behavior, 
the peak power used by the system lies below the bound given by the specification; 
or if for each behavior, the long-run average response time of the system
lies below the specified average response requirement.s
In~\cite{CDH08}, we showed that the quantitative language-inclusion problem
is PSPACE-complete for $\Max$-, $\LimSup$-, and $\LimInf$-automata, while
the decidability is unknown for (nondeterministic) limit-average and discounted-sum automata.
We also compared the expressive power of the different classes of quantitative languages
and showed that 
nondeterministic automata are strictly more expressive in the case of
limit-average and discounted-sum. 

In this paper, we investigate alternative ways of comparing the \emph{expressive power}
of weighted automata. First, we consider the cut-point languages of weighted
automata, a notion borrowed from the theory of probabilistic automata~\cite{Rabin63}.
Given a threshold $\eta \in \real$, the cut-point language of a quantitative
language $L$ is the set of all words $w$ with value $L(w) \geq \eta$, thus
a boolean language.
We show that deterministic limit-average and 
discounted-sum automata can define cut-point languages that are not 
$\omega$-regular. 
Note that there exist $\omega$-regular languages that cannot be expressed
as a cut-point language of a limit-average or discounted-sum automaton~\cite{CDH08}.
Then, we consider the special case where the threshold~$\eta$ is isolated, meaning
that there is no word with value in the neighborhood of~$\eta$. We argue that 
isolated cut-point languages have stability properties, by showing that they
remain unchanged under small perturbations of the transition weights. 
Furthermore, we show that every discounted-sum automaton with isolated cut-point defines
an $\omega$-regular language, and the same holds for deterministic limit-average
automata. This question is open for nondeterministic limit-average automata.
Finally, we consider a boolean counterpart of limit-average and discounted-sum 
automata in which all transitions have weight~$0$ or~$1$.
Of special interest is a proof that limit-average automata with rational weights in 
the interval $[0,1]$ can be reduced to automata with boolean weights.
Therefore, the restriction to boolean weights does not change the class of quantitative 
languages definable by limit-average automata; on the other hand, we show that it reduces that of
discounted-sum automata. 

In the second part of this paper, we study the \emph{closure properties} of quantitative languages.
It is natural and convenient to decompose a specification or a design 
into several components, and to apply composition operations
to obtain a complete specification. 
We consider a natural generalization of the classical operations
of union, intersection and complement of boolean languages. We define the \emph{maximum}, 
\emph{minimum}, and \emph{sum} of two quantitative languages $L_1$ and $L_2$ as the quantitative language
that assigns $\max(L_1(w),L_2(w))$, $\min(L_1(w),L_2(w))$, and $L_1(w) + L_2(w)$
to each word $w$. 
The \emph{complement}~$L^c$ of a quantitative language~$L$ is defined by $L^c(w) = 1-L(w)$
for all words $w$.\footnote{One can define $L^c(w) = k-L(w)$ for any constant $k$ without changing 
the results of this paper.}   
The sum is a natural way of composing two automata
if the weights represent costs ({\it e.g.}, energy consumption). 
We give other examples in Section~\ref{sec:definitions} to illustrate 
the composition operations and the use of quantitative languages 
as a specification framework.

We give a complete picture of the closure properties of 
the various classes of quantitative languages (over finite and infinite words) under maximum,
minimum, complement and sum (see \tablename~\ref{tab:closure-properties}).
For instance, limit-average automata are not closed under sum and complement,
while nondeterministic discounted-sum automata are closed under sum but not under complement.
All other classes of weighted automata are closed under sum.
For infinite words, the closure properties of $\Max$-, $\LimSup$-, and $\LimInf$-automata
are obtained as a direct extension of the results for the boolean finite automata,
while for $\LimAvg$- and $\Disc$-automata, the proofs respectively require the analysis of the structure
of the automata cycles and properties of the solutions of polynomials with rational coefficients.
Note that the quantitative language-inclusion problem 
``is $L_{A}(w)\leq L_{B}(w)$  for all words~$w$?'' reduces to 
closure under sum and complement because it is equivalent 
to the question of the non-existence of a word $w$ such that $L_{A}(w) + L^c_{B}(w) > 1$,
that is an \emph{emptiness} question which is decidable for all classes
of quantitative languages~\cite{CDH08}.
Also note that deterministic limit-average and discounted-sum automata are not 
closed under maximum, which implies 
that nondeterministic automata are strictly more expressive in these cases 
(because the maximum can be obtained by an initial nondeterministic choice). 
\smallskip\noindent{\it Related work.} Functions such as limit average (or mean payoff) and discounted 
sum have been studied extensively in the branching-time context of game theory~\cite{Sha53,EM79,Condon92,ZwickP96,CAHS03}.
It is therefore natural to use the same functions in the linear-time context of languages
and automata. 

Weighted automata with discounted sum have been considered in~\cite{DrosteR07},
with multiple discount factors and a boolean acceptance condition (Muller or B\"uchi); 
they are shown to be equivalent to a weighted monadic second-order logic with
discounting. Several other works have considered quantitative generalizations of 
languages, over finite words~\cite{DrosteGastin07}, over trees~\cite{DrosteKR08},
or using finite lattices~\cite{GurfinkelC03}, but none of these works
has addressed the expressiveness questions and closure properties for quantitative 
languages that are studied here.

The lattice automata of~\cite{LatticeAutomata07} map finite words to values from a finite
lattice. The lattice automata with B\"uchi condition are analogous to our 
$\LimSup$ automata, and their closure properties are established there.
However, the other classes of quantitative automata ($\Sum$, limit-average, 
discounted-sum) are not studied there as they cannot be defined using lattice 
operations and finite lattices.

\section{Quantitative Languages}\label{sec:definitions}

A \emph{quantitative language} $L$ over a finite alphabet $\Sigma$
is either a mapping $L: \Sigma^{+} \to \real$ or a mapping $L: \Sigma^{\omega} \to \real$, 
where $\real$ is the set of real numbers. 

\paragraph{\bf Weighted automata.}
A \emph{weighted automaton} is a tuple $A=\tuple{Q,q_I,\Sigma,\delta,\weight}$
where:
\begin{itemize}
\item $Q$ is a finite set of states, $q_I \in Q$ is the initial state, and $\Sigma$ is a finite alphabet;
\item $\delta \subseteq Q \times \Sigma \times Q$ is a finite set of labelled transitions.
We assume that $\delta$ is \emph{total}, that is for all 
$q \in Q$ and $\sigma \in \Sigma$, there exists $(q,\sigma,q') \in \delta$ 
for at least one $q' \in Q$;
\item $\weight: \delta \to \rat$ is a \emph{weight} function, where $\rat$ is the 
set of rational numbers. We assume that rational numbers are encoded as pairs of 
integers in binary.
\end{itemize}

We say that $A$ is \emph{deterministic} if for all 
$q \in Q$ and $\sigma \in \Sigma$, there exists $(q,\sigma,q') \in \delta$ 
for exactly one $q' \in Q$. We sometimes call automata \emph{nondeterministic}
to emphasize that they are not necessarily deterministic.
 
A \emph{run} of $A$ over a finite (resp. infinite) word $w=\sigma_1 \sigma_2 \dots$ 
is a finite (resp. infinite) sequence $r = q_0 \sigma_1 q_1 \sigma_2 \dots $ 
of states and letters such that 
\begin{compressEnum}
\itCompress $q_0 = q_I$, and
\itCompress $(q_i,\sigma_{i+1},q_{i+1}) \in \delta$ for all $0 \leq i < \abs{w}$.
\end{compressEnum}
We denote by $\weight(r) = v_0 v_1 \dots$ the sequence of weights that occur in~$r$ 
where $v_i = \weight(q_i,\sigma_{i+1},q_{i+1})$ for all $0 \leq i < \abs{w}$.

Given a \emph{value function} $\Val: \rat^+ \to \real$ (resp. $\Val: \rat^{\omega} \to \real$), 
we say that the $\Val$-automaton~$A$ defines the quantitative language $L_A$ such 
that for all $w \in \Sigma^{+}$ (resp. $w \in \Sigma^{\omega}$):
$$L_A(w) = \sup \{\Val(\weight(r)) \mid r \text{ is a run of } A \text{ over } w\}.$$

We consider the following value functions to define quantitative languages.
Given a finite sequence $v= v_1 \dots v_n$ of rational numbers, define
\begin{itemize}
\item $\Last(v) = v_n$;
\item $\Maxf(v) = \sup \{v_i \mid 1 \leq i \leq n\}$;
\item $\Sum(v) = \displaystyle\sum_{i=1}^{n} v_i$;
\end{itemize}
Given an infinite sequence $v=v_0 v_1 \dots$ of rational numbers, define
\begin{itemize}
\item $\Max(v)    = \sup \{v_n \mid n \geq 0\}$;
\item $\LimSup(v) = \displaystyle\limsup_{n\to\infty} \ v_n = \lim_{n\to\infty} \sup \{v_i \mid i \geq n\}$;
\item $\LimInf(v) = \displaystyle\liminf_{n\to\infty} \ v_n = \lim_{n\to\infty} \inf \{v_i \mid i \geq n\}$;
\item $\LimAvg(v) = \displaystyle\liminf_{n\to\infty} \ \frac{1}{n} \sum_{i=0}^{n-1} v_i$;
\item For $0 < \lambda < 1$, $\Disc_{\lambda}(v) = \displaystyle \sum_{i=0}^{\infty} \lambda^i \cdot v_i$;
\end{itemize}

Note that B\"uchi and coB\"uchi automata are special cases of respectively 
$\LimSup$- and $\LimInf$-automata, where all weights are either $0$ or $1$.


\paragraph{\bf Notations.} Classes of weighted automata over infinite words are denoted 
with acronyms of the form 
$xy$ where $x$ is either {\sc N}(ondeterministic), {\sc D}(eterministic), 
or \nd(when deterministic automata have the same expressiveness as nondeterministic automata),
and $y$ is one of the following: {\sc Sup}, \Ls(LimSup), \Li(LimInf), \La(LimAvg), or \Di.
For B\"uchi and coB\"uchi automata, we use the classical acronyms NBW, DBW, NCW, etc.

\paragraph{\bf Reducibility.} 
A class $\C$ of weighted automata is \emph{reducible}
to a class $\C'$ of weighted automata if for every $A \in \C$ there exists 
$A' \in \C'$ such that $L_A=L_{A'}$, i.e. $L_{A}(w)=L_{A'}(w)$ for all (finite or infinite) words $w$.
In particular, a class of weighted automata \emph{can be determinized}
if it is reducible to its deterministic counterpart.
Reducibility relationships for (non)deterministic weighted automata
are given in~\cite{CDH08}.

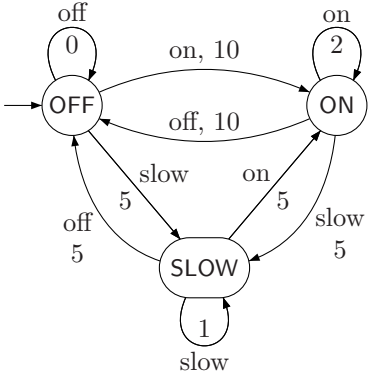
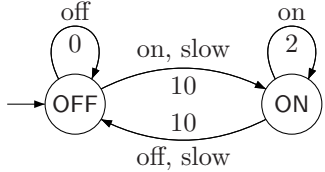
\begin{figure*}[!tb]
\begin{center}
     \subfigure[Limit-average automaton $A$. \label{fig:left}]{\unitlength=.8mm
\def\fsize{\normalsize}

\begin{picture}(105,63)(0,0)

{\fsize

\node[Nmarks=i](off)(30,45){{\small {\sf OFF}}}
\node[Nmarks=n](on)(74,45){{\small {\sf ON}}}
\node[Nmarks=n, Nw=15](slow)(52,18){{\small {\sf SLOW}}}

\drawloop[ELside=l, ELdist=1, loopCW=y, loopdiam=8, loopangle=90](off){off}
\drawloop[ELside=r, ELdist=1.5, loopCW=y, loopdiam=8, loopangle=90](off){$0$}
\drawedge[ELpos=50, ELside=l, ELdist=1, curvedepth=6](off,on){on, $10$}
\drawedge[ELpos=52, ELside=l, ELdist=1, syo=-1, curvedepth=0](off,slow){slow}
\drawedge[ELpos=50, ELside=r, ELdist=1, syo=-1, curvedepth=0](off,slow){$5$}

\drawloop[ELside=l, ELdist=1, loopCW=y, loopdiam=8, loopangle=90](on){on}
\drawloop[ELside=r, ELdist=1.5, loopCW=y, loopdiam=8, loopangle=90](on){$2$}
\drawedge[ELpos=50, ELside=r, ELdist=1, curvedepth=6](on,off){off, $10$}
\drawbpedge[ELpos=47, ELside=l, ELdist=-0.5](on,270,15,slow,0,15){\begin{tabular}{c}slow \\ $5$ \end{tabular}}

\drawloop[ELside=r, ELdist=1, loopCW=n, loopdiam=8, loopangle=-90](slow){slow}
\drawloop[ELside=l, ELdist=1.5, loopCW=n, loopdiam=8, loopangle=-90](slow){$1$}
\drawbpedge[ELpos=50, ELside=l, ELdist=0, exo=0](slow,180,15,off,270,15){\begin{tabular}{c}off \\ $5$ \end{tabular}}
\drawedge[ELpos=51, ELside=l, ELdist=1, eyo=-1, curvedepth=0](slow,on){on}
\drawedge[ELpos=50, ELside=r, ELdist=1, eyo=-1, curvedepth=0](slow,on){$5$}



}
\end{picture}}
     \subfigure[Limit-average automaton $B$. \label{fig:right}]{\unitlength=.8mm
\def\fsize{\normalsize}

\begin{picture}(105,63)(0,0)

{\fsize

\node[Nmarks=i](off)(34,30){{\small {\sf OFF}}}
\node[Nmarks=n](on)(70,30){{\small {\sf ON}}}

\drawloop[ELside=l, ELdist=1, loopCW=y, loopdiam=8, loopangle=90](off){off}
\drawloop[ELside=r, ELdist=1.5, loopCW=y, loopdiam=8, loopangle=90](off){$0$}
\drawedge[ELpos=50, ELside=l, ELdist=1, curvedepth=6](off,on){on, slow}
\drawedge[ELpos=50, ELside=r, ELdist=1.5, curvedepth=6](off,on){$10$}
\drawloop[ELside=l, ELdist=1, loopCW=y, loopdiam=8, loopangle=90](on){on}
\drawloop[ELside=r, ELdist=1.5, loopCW=y, loopdiam=8, loopangle=90](on){$2$}
\drawedge[ELpos=50, ELside=l, ELdist=1, curvedepth=6](on,off){off, slow}
\drawedge[ELpos=50, ELside=r, ELdist=1.5, curvedepth=6](on,off){$10$}



}
\end{picture}}
\end{center}
\caption{Specifications for the power consumption of a motor. 
$A$ refines $B$, i.e. $L_A \leq L_{B}$.\label{fig:motor-spec}}
\end{figure*}

\paragraph{\bf Composition.} 
Given two quantitative languages $L$ and $L'$ over $\Sigma$,
and a rational number $c$,
we denote by $\max(L,L')$ (resp. $\min(L,L')$, $L+L'$, $c+L$, and $cL$)
the quantitative language that assigns $\max\{L(w),L'(w)\}$
(resp. $\min\{L(w),L'(w)\}$, $L(w) + L'(w)$, $c+L(w)$, and $c\cdot L(w)$) to each word 
$w \in \Sigma^{+}$ (or $w \in \Sigma^{\omega}$). We say that $c+L$ is the \emph{shift by $c$}
of $L$ and that $cL$ is the \emph{scale by $c$} of $L$.
The language $1-L$ is called the \emph{complement} of $L$.
The $\max$, $\min$ and complement operators for quantitative languages 
generalize respectively the union, intersection and complement 
operator for boolean languages. For instance, De Morgan's laws hold
(the complement of the max of two languages is the min of their complement, etc.)
and complementing twice leave languages unchanged.



\smallskip\noindent{\it Example 1.} We consider a simple illustration of the use of limit-average automata
to model the power consumption of a motor. The automaton $B$ in \figurename~\ref{fig:right}
specifies the maximal power consumption to maintain the motor on or off, and the 
maximal consumption for a mode change. The specification abstracts away that a mode
change can occur smoothly with the $\mathit{slow}$ command. A refined specification $A$ is 
given in \figurename~\ref{fig:left} where the effect of slowing down is captured
by a third state. One can check that $L_{A}(w)\leq L_{B}(w)$ 
for all words~$w \in \{\mathit{on},\mathit{off},\mathit{slow}\}^\omega$.
Given two limit-average automata that model the power consumption of 
two different motors, the maximal, minimal, and the sum of average power 
consumption are obtained by composing the automata under max, min and sum 
operations, respectively. 

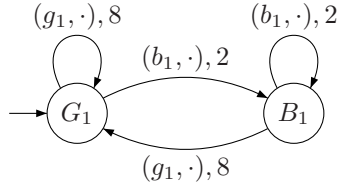
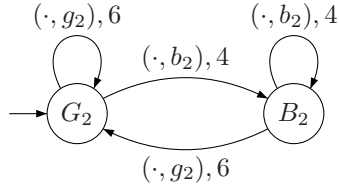
\begin{figure*}[!t]
\begin{center}
     \subfigure[100 dollars invested in bank $A_1$. \label{fig:bank1}]{\unitlength=.8mm
\def\fsize{\normalsize}

\begin{picture}(105,30)(0,0)

{\fsize

\node[Nmarks=i](good)(34,12){$G_1$}
\node[Nmarks=n](bad)(70,12){$B_1$}

\drawloop[ELside=l, ELdist=1, loopCW=y, loopdiam=8, loopangle=90](good){$(g_1,\cdot),8$}
\drawedge[ELpos=50, ELside=l, ELdist=1, curvedepth=6](good,bad){$(b_1,\cdot),2$}
\drawloop[ELside=l, ELdist=1, loopCW=y, loopdiam=8, loopangle=90](bad){$(b_1,\cdot),2$}
\drawedge[ELpos=50, ELside=l, ELdist=1, curvedepth=6](bad,good){$(g_1,\cdot),8$}



}
\end{picture}}
     \subfigure[100 dollars invested in bank $A_2$. \label{fig:bank2}]{\unitlength=.8mm
\def\fsize{\normalsize}

\begin{picture}(105,30)(0,0)

{\fsize

\node[Nmarks=i](good)(34,12){$G_2$}
\node[Nmarks=n](bad)(70,12){$B_2$}

\drawloop[ELside=l, ELdist=1, loopCW=y, loopdiam=8, loopangle=90](good){$(\cdot,g_2),6$}
\drawedge[ELpos=50, ELside=l, ELdist=1, curvedepth=6](good,bad){$(\cdot,b_2),4$}
\drawloop[ELside=l, ELdist=1, loopCW=y, loopdiam=8, loopangle=90](bad){$(\cdot,b_2),4$}
\drawedge[ELpos=50, ELside=l, ELdist=1, curvedepth=6](bad,good){$(\cdot,g_2),6$}



}
\end{picture}}
\end{center}
\caption{The discounted-sum automata models of two banks.\label{fig:bank-spec}}
\end{figure*}

\smallskip\noindent{\it Example 2.} Consider an investment of 100~dollars
that can be made in two banks~$A_1$ and~$A_2$ as follows: 
(a)~100~dollars to bank~$A_1$, (b)~100~dollars to bank $A_2$, or
(c)~50~dollars to bank~$A_1$ and 50~dollars to bank~$A_2$.
The banks can be either in a good state (denoted $G_1$, $G_2$)
or in a bad state (denoted $B_1$, $B_2$).
If it is in a good state, then~$A_1$ offers 8\% reward while~$A_2$ offers 6\% reward.
If it is in a bad state, then~$A_1$ offers 2\% reward while~$A_2$ offers 4\% reward.
The change of state is triggered by the input symbols~$b_1, b_2$ (from a good
to a bad state) and~$g_1, g_2$ (from a bad to a good state).
The rewards received earlier weight more than rewards received later 
due to inflation represented by the discount factor.
The automata~$A_1$ and~$A_2$ in Figure~\ref{fig:bank-spec} specify the behavior 
of the two banks for an investment of 100~dollars, where the input alphabet is 
$\set{g_1,b_1} \times \set{g_2,b_2}$ (where the notation $(g_1,\cdot)$ 
represents the two letters $(g_1,g_2)$ and $(g_1,b_2)$, and similarly for the other symbols). 
If 50~dollars are invested in each bank, then we obtain automata~$C_1$ and~$C_2$ 
from~$A_1$ and~$A_2$ where each reward is halved. 
The combined automaton is obtained as the composition of~$C_1$ and~$C_2$ 
under the sum operation.

\section{Expressiveness Results for Weighted Automata}

The expressive power of weighted automata can be compared by mean of the
reducibility relation, saying that a class $\C$ of weighted automata  is at least
as expressive as a class $\C'$ if every quantitative language definable
by some automaton in $\C$ is also definable by some automaton in $\C'$.
The comparison includes boolean languages, considering them as a special case of quantitative languages
of the form $L:\Sigma^{\omega} \to \{0,1\}$.
It was shown in~\cite{CDH08} that a wide variety of classes of quantitative languages
can be defined by the different types of weighted automata, depending on the
value function and whether they are deterministic or not. This contrasts with
the situation for boolean languages where most of the classes of automata define
$\omega$-regular languages.
In this section, we investigate alternative ways of comparing the expressive power
of weighted automata and of classical finite automata. 
First, we use the cut-point languages of weighted automata to compare with
the class of $\omega$-regular languages, and then we use weighted automata
with boolean weights, i.e. all transitions have weight $0$ or $1$, to compare 
with general weighted automata.

\subsection{Cut-point languages}

Let $L$ be a quantitative language over infinite words and let $\eta \in \real$ be a threshold.
The \emph{cut-point language} defined by $(L,\eta)$ is the (boolean) language
$$L^{\geq \eta} = \{w \in \Sigma^{\omega} \mid L(w) \geq \eta\}.$$
Cut-point languages for finite words are defined analogously. They
have been first defined for probabilistic automata~\cite{Rabin63},
then generalized to inverse image recognition for semiring automata over finite words~\cite{CortesM00}.
It is easy to see that the cut-point languages of $\Maxf$- and $\Last$-automata are regular,
those of $\Sum$-automata are context-free, and 
those of $\Max$-, $\LimSup$-, and $\LimInf$-automata are $\omega$-regular.

We show that the classes of cut-point languages definable by (non)deterministic limit-average and 
discounted-sum automata are incomparable with the $\omega$-regular languages. 
The result follows from Theorem~\ref{theo:cut-point-language}, 
and from~\cite[Theorems 13 and 14]{CDH08}.

\begin{theorem}\label{theo:cut-point-language}
There exists deterministic limit-average and discounted-sum automata
whose cut-point language is not $\omega$-regular.
\end{theorem}

\begin{myProof}
Consider the alphabet $\Sigma=\set{a,b}$, and consider the languages $L_1$
that assigns to each word its long-run average number of $a$'s,
and $L_2$ that assigns the discounted sum of $a$'s. Note that $L_1$
is definable by a deterministic limit-average automaton,
and $L_2$ by a deterministic discounted-sum automaton.
It was shown in~\cite{Cha-TCS} that the cut-point language 
$L_1^{\geq 1}$ is complete for the third level of the Borel hierarchy,
and therefore is not $\omega$-regular.
We show that $L_2^{\geq 1}$ is not $\omega$-regular.

Given a finite word $w \in \Sigma^*$, let $r_a(w) = \sum_{i\mid w_i = a} \lambda^i$
be the discounted sum of $a$'s in $w$. We say that $w$ is \emph{ambiguous}
if $1- \frac{\lambda^{\abs{w}}}{1-\lambda} \leq r_a(w) < 1$. The ambiguity lies
in that some continuations of $w$ (namely $w.a^{\omega}$) are in $L_2^{\geq 1}$ and some are not
(namely $w.b^{\omega}$).
We show that for all $\lambda > \frac{1}{2}$, if $w$ is ambiguous,
then either $w.a$ or $w.b$ is ambiguous, which entails 
that there exists an infinite word $w^{\preceq}$ all of whose 
finite prefixes are ambiguous (and $L_2(w^{\preceq}) = 1$).
To do this, assume that $1- \frac{\lambda^{\abs{w}}}{1-\lambda} \leq r_a(w) < 1$,
and show that either
$1- \frac{\lambda^{1+\abs{w}}}{1-\lambda} \leq r_a(w.a) < 1$
or $1- \frac{\lambda^{1+\abs{w}}}{1-\lambda} \leq r_a(w.b) < 1$.
Since $r_a(w.a) = r_a(w) + \lambda^{\abs{w}}$ and $r_a(w.b) = r_a(w)$,
we have to show that 
$1- \frac{\lambda^{\abs{w}}}{1-\lambda} \leq r_a(w) < 1 - \lambda^{\abs{w}}$
or $1- \frac{\lambda^{1+\abs{w}}}{1-\lambda} \leq r_a(w) < 1$.
This holds if $1- \frac{\lambda^{1+\abs{w}}}{1-\lambda} < 1 - \lambda^{\abs{w}}$,
which is equivalent to $\lambda > \frac{1}{2}$.

Now, we show that if there exists a nondeterministic B\"uchi automaton $A$ for $L_2^{\geq 1}$,
then the set of states $S_n$ reached in $A$ by reading the first $n$ letters
of $w^{\preceq}$ (which we denote by $w_{[1\dots n]}^{\preceq}$) should be different for each $n$, i.e. $n \neq m$ implies $S_n \neq S_m$.
Towards a contradiction, assume that $S_n = S_m$ for $n < m$.
Then for all $w' \in \Sigma^\omega$, we have $w_{[1\dots n]}^{\preceq}.w' \in L_2^{\geq 1}$ 
if and only if $w_{[1\dots m]}^{\preceq}.w' \in L_2^{\geq 1}$.
In particular, for $w' = w^{\preceq}_{[m+1\dots]}$, this shows that 
$L_2(w_{[1\dots n]}^{\preceq}.w') = 1 = L_2(w_{[1\dots m]}^{\preceq}.w')$
since $L_2(w^{\preceq}) = 1$ and $r_a(w_{[1\dots n]}) \leq r_a(w_{[1\dots m]})$.
This yields
$$r_a(w_{[1\dots n]}) + \lambda^n \cdot r_a(w') = 1 = r_a(w_{[1\dots m]}) + \lambda^m \cdot r_a(w')$$
that is, by eliminating $r_a(w')$, $\lambda^{m-n} (1-P(\lambda)) = 1 - Q(\lambda)$
where $P(\lambda)) = r_a(w_{[1\dots n]})$ and $Q(\lambda) = r_a(w_{[1\dots m]})$
are polynomials of respective degree $n-1$ and $m-1$, and with coefficients in 
the set $\{0,1\}$. First, observe that the equation is not identically $0$
because the coefficient of the term of degree $0$ is not $0$ (as the first letter
of $w^{\preceq}$ must be $b$ since $a$ is not ambiguous). Second, every
coefficient in the equation is in the set $\{-1,0,1,2\}$, and a classical
result shows that if $\frac{p}{q}$ is a solution of a polynomial equation
with $p$ and $q$ mutually prime, then $p$ divides the coefficient of degree $0$,
and $q$ divides the coefficient of highest degree. Therefore, no rational number 
in the interval $]\frac{1}{2},1[\,$ can be a solution.
This shows that $n \neq m$ implies $S_n \neq S_m$, and the automaton $A$ cannot have
finitely many states.
\end{myProof}


We note that cut-point languages are not stable under arbitrarily small 
perturbations of the transition weights, nor of the value of the cut-point. 
Consider the quantitative languages $L_1$, $L_2$ from the proof of Theorem~\ref{theo:cut-point-language}.
If for instance a limit-average automaton $A$ assigns weight $1+\epsilon$ 
to the $a$'s and $0$ to the $b$'s, its cut-point language $L_A^{\geq 1}$
is clearly not different from $L_1^{\geq 1}$ that assigns to each word its long-run 
average number of $a$'s, no matter the value of $\epsilon > 0$. The same
holds with respect to $L_2$ if $A$ is interpreted as a discounted-sum automaton.

In the theory of probabilistic automata, where finite words are assigned a probability
of acceptance, the cut-point languages may also be non-regular. Therefore, 
one considers the special case where the cut-point is isolated, and
shows that the cut-point languages are then regular~\cite{Rabin63}.

A number $\eta$ is an \emph{isolated cut-point} of a quantitative language $L$
if there exists $\epsilon > 0$ such that
$$\abs{L(w) - \eta} > \epsilon \text{ for all } w \in \Sigma^{\omega}.$$

We show that every discounted-sum automaton with isolated cut-point defines
an $\omega$-regular language, and that this also holds for deterministic limit-average
automata. We also argue that this notion has stability properties, in that 
isolated cut-point languages remain unchanged under small perturbations of 
the transition weights. This follows from a more general result about
the robustness of weighted automata.

A class of weighted automata is robust if a small (syntactical) perturbation
in the weights of an automaton induces only a small (semantical) perturbation 
in the values of the words in the quantitative language of the automaton,
and the semantical perturbation tends to $0$ when the syntactical perturbation tends to $0$.
To formally define robustness, we need $\epsilon$-approximations of automata,
and distance between quantitative languages.

Let $A=\tuple{Q,q_I,\Sigma,\delta,\weight}$ be a (nondeterministic) weighted automaton,
and let $\epsilon \in \posreal$. We say that a weighted automaton $B=\tuple{Q',q'_I,\Sigma,\delta',\weight'}$
is an \emph{$\epsilon$-approximation} of $A$ if
\begin{itemize}
\item $Q' = Q$, $q'_I = q_I$, $\delta' = \delta$, and
\item $\abs{\weight'(q,\sigma,q') - \weight(q,\sigma,q')} \leq \epsilon$ for all $(q,\sigma,q') \in \delta$.
\end{itemize}
The \emph{$\sup$-distance} between two quantitative languages $L_1,L_2: \Sigma^{\omega} \to \real$
is defined by 
$$D_{\sup}(L_1,L_2) = \sup_{w \in \Sigma^{\omega}} \abs{L_1(w) - L_2(w)}.$$

We say that a class $\C$ of weighted automata is \emph{uniformly robust} if for all $\eta \in \sposreal$,
there exists $\epsilon \in \sposreal$ such that for all automata $A,B \in \C$ where 
$B$ is an $\epsilon$-approximation of $A$, we have $D_{\sup}(L_A,L_B) \leq \eta$.
Note that uniform robustness implies a weaker notion of robustness where
a class $\C$ of weighted automata is called \emph{robust} if for all automata $A \in \C$
and for all $\eta \in \sposreal$, there exists $\epsilon \in \sposreal$ such that 
for all $\epsilon$-approximation $B$ of $A$ (with $B \in \C$), we have $D_{\sup}(L_A,L_B) \leq \eta$.

\begin{theorem}\label{theo:uniformly-robust}
The classes of (non)deterministic $\Max$-, $\LimSup$-, $\LimInf$-, $\LimAvg$- and $\Disc$-automata
are uniformly robust.
\end{theorem}

\begin{myProof}
Let $A,B$ be two weighted automata with $B$ an $\epsilon$-approximation of $A$.
It is easy to see that for $\Max$-, $\LimSup$-, $\LimInf$- and $\LimAvg$-automata, 
the value of a run $r$ of $B$ differs by at most $\epsilon$ from the value of 
the same run in $A$. Therefore, $D_{\sup}(L_A,L_B) \leq \epsilon$ and we can take 
$\epsilon = \eta$. 
For $\Disc$-automata, the value of a run of $B$ differs by at most $\frac{\epsilon}{1-\lambda}$ 
from the value of the same run in $A$, where $\lambda$ is the discount factor. 
Therefore, we can take $\epsilon = \eta (1 - \lambda)$. 
\end{myProof}

As a corollary of Theorem~\ref{theo:uniformly-robust}, for an isolated cut-point $\eta$, 
the cut-point language $L^{\geq \eta}$ remains unchanged under small
perturbations of the transition weights. 

\begin{theorem}
Let $L_A$ be the quantitative language defined by a weighted automaton $A$,
and let $\eta$ be an isolated cut-point of $L_A$. There exists $\epsilon > 0$
such that for all $\epsilon$-approximations $B$ of $A$, $L_A^{\geq \eta} = L_B^{\geq \eta}$ 
(where $L_B$ is the quantitative language defined by $B$).
\end{theorem}

Now, we show that the isolated cut-point languages of 
deterministic discounted-sum and limit-average automata are $\omega$-regular. 
For nondeterministic automata, the same property holds in the discounted-sum case, 
but the question is open for limit average.

\begin{theorem}
Let $L$ be the quantitative language defined by a $\Disc$-automaton.
If $\eta$ is an isolated cut-point of $L$, then the cut-point language $L^{\geq \eta}$
is $\omega$-regular.
\end{theorem}

\begin{myProof}
Let $\lambda$ be the discount factor of the $\Disc$-automaton that defines $L$.
Since, $\eta$ is an isolated cut-point of $L$, let $\epsilon > 0$ such that
$\abs{L(w) - \eta} > \epsilon$ for all $w \in \Sigma^{\omega}$.
Let $n \in \nat$ such that $u_n = \frac{V\cdot \lambda^n}{1-\lambda} < \epsilon $
where $V = \max_{(q,\sigma,q')} \abs{\delta(q,\sigma,q')}$ is largest weight in $A$.
Consider any run $r$ in $A$ of length $n$, and let $\weight(r)$ be the $\lambda$-discounted
sum of the weights along $r$. Then, it should be clear that 
$\weight(r) \not\in [\eta - \epsilon + u_n, \eta + \epsilon - u_n]$, because
otherwise, the value of any (infinite) continuation of $r$ would lie in the
interval $[\eta - \epsilon, \eta + \epsilon]$, which would be a contradiction.
Moreover, if $\weight(r) \leq \eta - \epsilon + u_n$, then any (infinite) continuation of $r$
has value less than $\eta$, while if $\weight(r) \geq  \eta + \epsilon - u_n$,
then any (infinite) continuation of $r$ has value greater than $\eta$.
Therefore, the cut-point language $L^{\geq \eta}$ can be defined by the
unfolding up to length $n$ of the $\Disc$-automaton that defines $L$, 
in which the states that are reached via a path with value at least
$\eta + \epsilon - u_n$ are declared to be accepting, and have a self-loop on $\Sigma$.
\end{myProof}

\begin{theorem}
Let $L$ be the quantitative language defined by a deterministic $\LimAvg$-automaton.
If $\eta$ is an isolated cut-point of $L$, then the cut-point language $L^{\geq \eta}$
is $\omega$-regular.
\end{theorem}

\begin{myProof}
Let $A$ be a deterministic $\LimAvg$-automaton, defining the language $L$.
Consider the SCC-decomposition $C_1,C_2,\dots,C_k$ of the underlying graph of $A$.
For each $1 \leq i \leq k$, let $m_i$ and $M_i$ be the minimal and maximal
average weight of a cycle in $C_i$ (those values can be computed with
Karp's algorithm~\cite{Karp78}). 
It is easy to see that for every $1 \leq i \leq k$, for every $v \in [m_i, M_i]$, 
there exists a word $w \in \Sigma^{\omega}$ such that $L(w) = v$.
Therefore, since $\eta$ is an isolated cut-point of $L$, we have $\eta \not\in [m_i, M_i]$
for all $1 \leq i \leq k$. 
A DBW for $L^{\geq \eta}$ is obtained from $A$ by declaring to be accepting all 
states $q$ of $A$ such that $q \in C_i$ and $m_i > \eta$.
\end{myProof}


\subsection{Boolean weights}

We consider weighted automata with boolean set of weights, i.e. all transitions have weight $0$ or $1$.
The aim is to have a boolean counterpart to limit-average and discounted-sum
automata, and check if this changes their expressive power.
We show that the restriction does not change the class of quantitative 
languages definable by limit-average automata, but does reduce that of
discounted-sum automata.

Given a set $R \subseteq \real$, and a class $\C$ of nondeterministic weighted automata,
we denote by $\C_R$ the class of all automata in $\C$ whose weights are rational numbers in~$R$.


\begin{theorem}
The class of nondeterministic (resp. deterministic) $\LimAvg$-automata with weights in $[0,1] \cap \rat$
is reducible to the class of nondeterministic (resp. deterministic) $\LimAvg$-automata with weights $0$ and $1$ only.
\end{theorem}


\begin{myProof}
Given a \nla${}_{[0,1]}$-automaton $A=\tuple{Q,q_I,\Sigma,\delta,\weight}$,
we construct a \nla${}_{\{0,1\}}$-automaton $B$ such that $L_A = L_B$. 

First, let $W = \{\weight(q,\sigma,q') \mid (q,\sigma,q') \in \delta\}$
be the set of weights that occur in $A$, and let $n_A$ be the smallest integer $n$
such that for all $v \in W$, there exists $e \in \nat$ such that $v = \frac{e}{n}$
(i.e., $\frac{1}{n_A}$ is the greatest common divisor of the weights of $A$).
We define $B=\tuple{Q',q'_I,\Sigma,\delta',\weight'}$ as follows:
\begin{itemize}
\item $Q' = Q \times [n_A]$ (where $[n_A]$ denotes the set $\{0,1,\dots,n_A-1\}$).
Intuitively, when we reach a state $(q,i)$ in $B$, it means that the state $q$
was reachable in $A$ and that the sum of the weights to reach $q$ is of the form
$k + \frac{i}{n_A}$ for some integer $k$. In $B$ however, the sum of the weights
to reach $(q,i)$ will then be $k$, and we store in the discrete state the information
that the remainder weight is $\frac{i}{n_A}$. Whenever this remainder exceeds $1$,
we introduce a weight $1$ and decrement the remainder.
\item $q'_I = (q_I,0)$;
\item for each transition $(q,\sigma,q') \in \delta$ and each value $i \in [n_A]$,
the following transitions are in $\delta'$ (where $v = \weight(q,\sigma,q')$):
	\begin{itemize}
	\item $((q,i),\sigma,(q',j))$ for $j=i+(v-1)\cdot n_A$ if $\frac{i}{n_A} + v \geq 1$; 
		the weight of such a transition is $1$ in $\weight'$,
	\item $((q,i),\sigma,(q',j))$ for $j=i+v\cdot n_A$ if $\frac{i}{n_A} + v < 1$; 
		the weight of such a transition is $0$ in $\weight'$.
	\end{itemize}
Note that in the above, $v\cdot n_A$ is an integer and $j \in [n_A]$.
\end{itemize}

There is a straightforward correspondence between the runs in $A$ and the runs in $B$.
Moreover, if the average weight of a prefix of length $n$ of a run in $A$ is $\frac{S}{n}$,
then the average weight of the prefix of length $n$ of the corresponding run in $B$
is between $\frac{S}{n}$ and $\frac{S+1}{n}$. Hence the difference tends to $0$
when $n \to \infty$. Therefore, the value of a run in $A$ is the same as the
value of the corresponding run in $B$, and therefore $L_A = L_B$.

Finally, note that if $A$ is deterministic, then $B$ is deterministic.
\end{myProof}

\begin{theorem}
The class of deterministic $\Disc$-automata with rational weights in $[0,1]$
is not reducible to the class of (even nondeterministic) $\Disc$-automata with weights $0$ and $1$ only.
\end{theorem}



\begin{myProof}
Given a discount factor $0 < \lambda <1$, consider the \ndi${}_{[0,1]}$ over $\Sigma = \{a,b\}$ 
that consists of a single state with a self-loop over $a$ with weight $\frac{1+\lambda}{2}$ 
and a self-loop over $b$ with weight $0$. 
Let $L_{\lambda}$ be the quantitative language defined by this automaton.
Towards a contradiction, assume that this language is defined by a \ndi${}_{\{0,1\}}$ $A$.
First, consider the word $a b^{\omega}$ whose value in $L_{\lambda}$ is $\frac{1+\lambda}{2} < 1$. 
This entails that $A$ cannot have a transition from the initial state over $a$ with weight $1$ 
(as this would imply that $L_A(a b^{\omega}) \geq 1$).
Now, the maximal value that $L_A$ can assign to the word $a^{\omega}$ is 
$\lambda + \lambda^2 + \lambda^3 + \cdots = \frac{\lambda}{1-\lambda}$ which is
strictly smaller than $L_{\lambda}(a^{\omega}) = \frac{1+\lambda}{2(1-\lambda)}$.
This shows that $A$ cannot exist.
\end{myProof}

\section{The Closure Properties of Weighted Automata}\label{sec:closure-properties}

We study the closure properties of weighted automata
with respect to $\max$, $\min$, complement and sum. We say that a class $\C$ of weighted
automata is \emph{closed} under a binary operator $\op(\cdot,\cdot)$ 
(resp. a unary operator $\op'(\cdot)$) if for all $A_1,A_2 \in \C$,
there exists $A_{12} \in \C$ such that $L_{A_{12}} = \op(L_{A_1},L_{A_2})$
(resp. $L_{A_{12}} = \op'(L_{A_1})$).
All closure properties that we present in this paper are constructive:
when $\C$ is closed under an operator,
we can always construct the automaton $A_{12} \in \C$ given $A_1,A_2 \in \C$.
We say that the \emph{cost} of the closure property of $\C$ under a binary operator $\op$ 
is at most $O(f(n_1,m_1,n_2,m_2))$ if 
for all automata $A_1,A_2 \in \C$ with $n_i$ states and $m_i$ transitions (for $i=1,2$ respectively),
the constructed automaton $A_{12} \in \C$ such that $L_{A_{12}} = \op(L_{A_1},L_{A_2})$
has at most $O(f(n_1,m_1,n_2,m_2))$ many states.
Analogously, the \emph{cost} of the closure property of $\C$ under a unary operator $\op'$ 
is at most $O(f(n,m))$ if 
for all automata $A_1 \in \C$ with $n$ states and $m$ transitions,
the constructed automaton $A_{12} \in \C$ such that $L_{A_{12}} = \op'(L_{A_1})$
has at most $O(f(n,m))$ many states.
For all reductions presented, the size of the largest weight in $A_{12}$ 
is linear in the size~$p$ of the largest weight in~$A_1,A_2$ (however, 
the time needed to compute the weights is quadratic in $p$,
as we need addition, multiplication, or comparison, which are quadratic operations
over the rational numbers).


Notice that every class of weighted automata is closed under shift by~$c$ and
under scale by~$\abs{c}$ for all~$c \in \rat$. For $\Sum$-automata
and discounted-sum automata, we can define the shift by~$c$ by making a copy of the initial
states and adding $c$ to the weights of all its outgoing transitions.
For the other automata, it suffices to add~$c$
to (resp. multiply by~$\abs{c}$) all weights of an automaton to obtain the
automaton for the shift by~$c$ (resp. scale by~$\abs{c}$) of its language. 
Therefore, all closure properties also hold if the complement of a 
quantitative language~$L$ was defined as~$k-L$ for any constant~$k$.

Our purpose is the study of quantitative languages over infinite words.
For the sake of completeness, we first give an overview of the closure 
properties for finite words.

\subsection{Closure properties for finite words}

We successively consider closure under $\max$, $\min$, complement,
and sum for weighted automata over finite words.
\tablename~\ref{tab:closure-properties}(a)
summarizes the closure properties of $\Maxf$-, $\Last$- and $\Sum$-automata.

\begin{theorem}\label{theo:max-closure-finite}
Deterministic $\Max$- and $\Last$-automata are closed under $\max$,
with cost $O(n_1 \cdot n_2)$.
Nondeterministic $\Max$-, $\Last$- and $\Sum$-automata are closed under $\max$,
with cost $O(n_1 + n_2)$.
Deterministic $\Sum$-automata are not closed under $\max$.
\end{theorem}

\begin{myProof}
For the nondeterministic automata, the result follows from the fact that 
the $\max$ operator can be obtained by an initial nondeterministic choice 
between two quantitative automata. For deterministic $\Max$- and $\Last$-automata,
the result follows from the fact that the classes of nondeterministic $\Max$- and $\Last$-automata are
reducible\footnote{We say that a class $\C$ of quantitative 
automata is \emph{reducible} to a class $\C'$ of quantitative automata if 
for every $A \in \C$ there exists $A' \in \C'$such that $L_A=L_{A'}$.} to
their respective deterministic counterpart. 
Finally, deterministic $\Sum$-automata are not closed under the $\max$ operator 
because the language over $\Sigma = \{a,b\}$ that assigns to each finite word $w \in \Sigma^{+}$ 
the number $\max\{L_a(w),L_b(w)\}$ where $L_{\sigma}(w)$ is the number of occurrences of $\sigma$ in $w$ (for $\sigma = a,b$)
is definable by the max of two deterministic-$\Sum$ languages, but not 
by a deterministic $\Sum$-automaton (Theorem~4 in~\cite{CDH08}).
\end{myProof}

\begin{theorem}\label{theo:min-closure-finite}
Deterministic and nondeterministic $\Max$-automata
are closed under $\min$, with cost $O(n_1\cdot m_1 \cdot n_2 \cdot m_2)$.
Deterministic and nondeterministic $\Last$-automata
are closed under $\min$, with cost $O(n_1 \cdot n_2)$.
Deterministic and nondeterministic $\Sum$-automata are not closed
under $\min$.
\end{theorem}

\begin{myProof}
Given two $\Last$-automata $A_1$ and $A_2$ (over the same alphabet), 
we use the classical synchronized product $A_{12} = A_1 \times A_2$, 
where the weight of a transition in $A_{12}$ is the minimum of the corresponding
transition weights in $A_1$ and $A_2$. It is easy to see that $L_{A_{12}} = \min(L_{A_1}, L_{A_2})$.
If $A_1$ and $A_2$ are deterministic, then so is $A_{12}$.

The construction for $\Max$-automata is the same as for $\Max$-automata
over infinite words given in the proof of \theoremname~\ref{theo:max-closed-under-min}.

Finally, for $\Sum$-automata, consider the language $L_m$ over $\Sigma = \{a,b\}$ 
that assigns to each finite word $w \in \Sigma^{+}$ the value $\min\{L_a(w),L_b(w)\}$ 
where $L_{\sigma}(w)$ is the number of occurrences of $\sigma$ in $w$ (for $\sigma = a,b$).
We claim that $L_m$ is not definable by a nondeterministic $\Sum$-automaton.
Indeed, assume that the $\Sum$-automaton $A$ defines $L_m$.
First, every the sum of weights in every reachable cycle of $A$ over $a$'s 
must be at most $0$. Otherwise, we can reach the cycle with a finite word $w_1$
and obtain an arbitrarily large value for the word $w_1 a^i$ for $i$ sufficiently 
large, while for such $i$ the value of $w_1 a^i$ is the number of $b$'s
in $w_1$ which is independent of $i$. Analogously, the sum of weights in 
every reachable cycle of $A$ over $b$'s must be at most $0$.
Now, let $\beta= \max_{e \in \delta} \weight(e)$ be the maximal 
weight in $A$, and consider the word $w = a^n b^n$ for $n > 2\beta \cdot \abs{Q}$.
Every run of $A$ over $a^n$ (or over $b^n$) can be decomposed in possibly nested cycles 
(since $A$ is nondeterministic) and a remaining non-cyclic path of length
at most $\abs{Q}$. Hence, the value of any run over $w$ is at most $2\beta \cdot \abs{Q}$.
However, the value of $w$ should be $n$, yielding a contradiction.
\end{myProof}

\begin{theorem}\label{theo:closure-under-complement-finite}
Deterministic $\Last$- and $\Sum$-automata are closed under complement, 
with cost $O(n)$.
Nondeterministic $\Last$-automata are closed under complement, 
with cost $O(2^n)$.
Nondeterministic $\Sum$ automata, and both deterministic and nondeterministic 
$\Max$-automata are not closed under complement.
\end{theorem}

\begin{myProof}
To define the complement of the language of a deterministic $\Sum$ (or $\Last$-) automaton, 
it suffices to multiply all the weights by $-1$, and then shift the language by $1$.
For the class of nondeterministic $\Last$-automata, the result follows from the fact that 
it is reducible to its deterministic counterpart.

The negative result for $\Max$-automata follows from an analogous in the boolean
case (consider the language $L$ over $\{a,b\}$ such that $L(a^i) = 0$
for all $i \geq 1$, and $L(w) = 1$ for all words containing the letter $b$). 
Finally, according to the proof of \theoremname~\ref{theo:min-closure-finite}, 
the language $\min(L_a,L_b)$ where $L_{\sigma}(w)$ is the number of occurrences 
of $\sigma$ in $w$ (for $\sigma = a,b$) is not definable by a nondeterministic 
$\Sum$-automaton. Since $\min(L_a,L_b) = -\max(-L_a,-L_b)$ and 
\begin{compressEnum}
\itCompress $-L_a$ and $-L_b$ are definable by $\Sum$-automata, and
\itCompress nondeterministic $\Sum$-automata are closed under $\max$ (\theoremname~\ref{theo:max-closure-finite}),
\end{compressEnum}
the language $\max(-L_a,-L_b)$ is definable by a nondeterministic $\Sum$-automaton,
and the result follows.
\end{myProof}

\begin{theorem}\label{theo:closure-under-sum-finite}
Every class of weighted automata over finite words are closed under sum.
The cost is $O(n_1\cdot n_2)$ for $\Last$- and $\Sum$-automata,
and $O(n_1\cdot m_1 \cdot n_2 \cdot m_2)$ for $\Max$-automata.
\end{theorem}

\begin{myProof}
It is easy to see that the synchronized product of two $\Last$-automata (resp. $\Sum$-automata)
defines the sum of their languages if the weight of a joint transition
is defined as the sum of the weights of the corresponding transitions in the two 
$\Last$-automata (resp. $\Sum$-automata).

The construction for $\Max$-automata is the same as for $\Max$-automata
over infinite words given in the proof of \theoremname~\ref{theo:max-closed-under-sum}.
\end{myProof}

\begin{table}
\begin{center}
\begin{tabular}{|l|*{4}{c|}}
\hline
       & max. & min. & comp. & sum \\
\hline
$\Max$   &  \ok & \ok  & \ko   & \ok \\
\hline
$\Last$  &  \ok & \ok  & \ok   & \ok \\
\hline
Det. $\Sum$   &  \ko & \ok  & \ok   & \ok \\
\hline
Nondet.  $\Sum$   &  \ok & \ko  & \ko   & \ok \\
\hline
\multicolumn{5}{c}{(a) Finite words}
\end{tabular}\hfill
\begin{tabular}{|l|*{4}{c|}}
\hline
       & max. & min. & comp. & sum \\
\hline
\ndmax &  \ok & \ok  & \ko   & \ok \\
\hline
\ndli  &  \ok & \ok  & \ko   & \ok \\
\hline
\dls   &  \ok & \ok  & \ko   & \ok \\
\hline
\nls   &  \ok & \ok  & \ok   & \ok \\
\hline
\dla   &  \ko & \ko  & \ko   & \ko \\
\hline
\nla   &  \ok & \ko  & \ko   & \ko \\
\hline
\ddi   &  \ko & \ko  & \ok   & \ok \\
\hline
\ndi   &  \ok & \ko  & \ko   & \ok \\
\hline
\multicolumn{5}{c}{(b) Infinite words}
\end{tabular}\hfill
\end{center}
\caption{Closure properties.\label{tab:closure-properties}}
\end{table}

\subsection{Closure under $\max$ for infinite words}

The maximum of two quantitative languages defined by nondeterministic automata
can be obtained by an initial nondeterministic choice between the two automata.
This observation was also made in~\cite{DrosteR07} for discounted-sum automata.
For deterministic automata, a synchronized product can be used
for $\Max$ and $\LimSup$, while for $\LimInf$ we use the fact that \nli\/ 
is determinizable with an exponential blow-up~\cite{CDH08}.


\begin{theorem}\label{theo:max-closure}
The nondeterministic $\Max$-, $\LimSup$-, $\LimInf$-, $\LimAvg$- and $\Disc$-automata
are closed under $\max$, with cost $O(n_1+n_2)$,
the deterministic $\Max$- and $\LimSup$-automata with cost $O(n_1 \cdot n_2)$,
the deterministic $\LimInf$-automata with cost $O((m_1+m_2)\cdot 2^{n_1 + n_2})$.
\end{theorem}

\begin{myProof}[Sketch]
For all the nondeterministic quantitative automata, the result 
follows from the fact that the $\max$ operator can be achieved
with an initial nondeterministic choice between two quantitative 
automata.
For \dli, the result follows from the reducibility of \nli\/ to \dli\/
with an exponential blow-up~\cite{CDH08}.
We now prove that \dls\/ and \dmax are closed under $\max$ with cost $O(n_1 \cdot n_2)$.
Given two \dls\/ (or \dmax) $A_1$ and $A_2$ over the same alphabet, we construct the usual
synchronized product $A_{12} = A_1 \times A_2$, where the weight of a
transition in $A_{12}$ is the maximum of the corresponding
transition weights in $A_1$ and $A_2$.
It is easy to see that $L_{A_{12}} = \max(L_{A_1}, L_{A_2})$ in both cases .
\end{myProof}


\begin{theorem}
The deterministic $\LimAvg$- and $\Disc$-automata are not closed under $\max$.
\end{theorem}

\begin{myProof} 
The fact that \ddi\/ is not closed under $\max$ follows
from the proof of \theoremname~34 in~\cite{CDH08}, where it is shown that the quantitative language
$\max(L_1,L_2)$ cannot be defined by a \ddi, where $L_1$ (resp. $L_2$) is the language
defined by the \ddi\/ that assigns weight $1$ (resp. $0$) to $a$'s 
and weight $0$ (resp. $1$) to $b$'s. 

We now show that \dla\/ is not closed under $\max$.
Consider the alphabet $\Sigma=\set{a,b}$ and the quantitative languages
$L_a$ and $L_b$ that assign the value of long-run average
number of $a$'s and $b$'s, respectively.
There exists \dla\/ for $L_a$ and $L_b$.
We show that $L_m=\max(L_a,L_b)$ cannot be expressed by 
a \dla. By contradiction, assume that $A$ is a \dla\/
with set of states $Q$ that defines $L_m$.
Consider any reachable cycle $C$ over $a$'s in $A$. The sum of the weights 
of the cycle must be its length $\abs{C}$, as if we consider the 
word $w^*=w_C \cdot (a^{\abs{C}})^\omega$ where $w_C$
is a finite word whose run reaches $C$, the value
of $w^*$ in $L_m$ is $1$. It follows that the sum of the weights
of the cycle $C$ must be $\abs{C}$. Hence, the sum of the weights
of all the reachable cycles $C$ over $a$'s in $A$ is $\abs{C}$.

\newcommand{\wh}{\widehat}

Consider the infinite word $w_\infty=(a^{\abs{Q}} \cdot b^{2\abs{Q}})^\omega$,
and let $w_j=(a^{\abs{Q}} \cdot b^{2\abs{Q}})^j$.
Since $L_m(w_\infty)=\frac{2}{3}$, the run of $A$ over $w_\infty$ 
has value $\frac{2}{3}$.
It follows that for all $\varepsilon>0$, there is an integer $j_\varepsilon$, such that
for all $j\geq j_\varepsilon$, we have 
\[
\frac{\weight(w_j)}{\abs{w_j}} \geq \frac{2}{3} -\varepsilon
\]
where $\weight(w_j)$ is the sum of the weights of the run of $A$ over $w_j$.
Consider a word $\wh{w}_\infty$ constructed as follows. We start 
with the empty word $\wh{w}_0$ and the initial state $q_0$ of $A$, and for all $j\geq 0$, we construct 
$(\wh{w}_{j+1}, q_{j+1})$ from $(\wh{w}_j,q_j)$ as follows: the state $q_{j+1}$
is the last state of the run of $A$ from $q_j$ over $a^{\abs{Q}} \cdot b^{2\abs{Q}}$. 
This run has to contain a cycle $C_{j+1}$ over $a$'s. We set 
$\wh{w}_{j+1} = \wh{w}_j \cdot a^{\abs{Q} + \abs{C_{j+1}}} \cdot b^{2\abs{Q}}$.
Observe that for all $j \geq 1$, the run of $A$ over $w_\infty$
in the segment between $w_j$ and $w_{j+1}$ is identical to
the run from $q_j$ to $q_{j+1}$ up to the repetition of the cycle $C_{j+1}$
once more.
The word $\wh{w}_\infty$ is the limit of this construction ($\wh{w}_j$ is
a prefix of $\wh{w}_\infty$ for all $j \geq 0$).
Let $\alpha_j=\sum_{i=1}^j \abs{C_i}$.    
Since $1\leq \abs{C_i} \leq \abs{Q}$ we have $j \leq \alpha_j \leq j \cdot \abs{Q}$.
Hence we have the following equality:
$\frac{\weight(\wh{w}_j)}{\abs{\wh{w}_j}} 
=\frac{\weight(w_j) + \alpha_j}{\abs{w_j}+ \alpha_j}$.
Hence for all $\varepsilon>0$, there exists $j_\varepsilon$ such that 
for all $j \geq j_\varepsilon$ we have
\[
\begin{array}{rcl}
\displaystyle
\frac{\weight(\wh{w}_j)}{\abs{\wh{w}_j}} 
& \geq  & 
\displaystyle 
\frac{ \frac{2}{3}\cdot \abs{w_j}   - \varepsilon \cdot \abs{w_j} + \alpha_j }{ \abs
{w_j} + \alpha_j} \\[2ex]
& \geq & 
\displaystyle 
\frac{2}{3} -\varepsilon + \frac{1}{3} \cdot \frac{\alpha_j}{\abs{w_j} + \alpha_j} 
\\[1ex]
& \geq & 
\displaystyle 
\frac{2}{3} -\varepsilon + \frac{1}{3} \cdot \frac{j}{j \cdot(3 \abs{Q} + \abs
{Q})} \\[1ex]
& \geq & 
\displaystyle 
\frac{2}{3} -\varepsilon + \frac{1}{12 \abs{Q}} \\[1ex]
\end{array}
\] 
Hence we have $L_A(\wh{w}_\infty) \geq \frac{2}{3} + \frac{1}{12 \abs{Q}}$.
Since $1 \leq |C_i| \leq \abs{Q}$ for all $i\geq 1$, we have $L_m(\wh{w}_\infty)\leq\frac{2}{3}$
which is a contradiction.
\end{myProof}

\subsection{Closure under $\min$ for infinite words}

The next theorems generalize the closure property under intersection
of the boolean languages. The construction of the automaton for the 
$\min$ is a direct extension of the well-known constructions in the 
boolean case.

\begin{theorem}\label{theo:max-closed-under-min}
The (non)deterministic $\Max$-automata are closed under $\min$, with cost $O(n_1\cdot m_1 \cdot n_2 \cdot m_2)$,
\end{theorem}

\begin{myProof}
Let $A_1=\tuple{Q_1,q_I^1,\Sigma,\delta_1,\weight_1}$ and $A_2=\tuple{Q_2,q_I^2,\Sigma,\delta_2,\weight_2}$
be two \nmax. We construct a \nmax\/ $A_{12}=\tuple{Q,q_I,\Sigma,\delta,\weight}$ such that
$L_{A_{12}} = \min\{L_{A_1},L_{A_2}\}$. 
Let $V_i = \{\weight_i(e) \mid e \in \delta_i\}$ be the set of weights that appear in $A_i$ (for $i=1,2$),
and define:
\begin{itemize}
\item $Q = Q_1 \times V_1 \times Q_2 \times V_2$. Intuitively, we remember in 
a state $(q_1, v_1, q_2, v_2)$ the largest weights $v_1,v_2$ seen so far in the corresponding
runs of $A_1$ and $A_2$;
\item $q_I = (q_I^1, v_{\min}^1, q_I^2, v_{\min}^2)$ where $v_{\min}^i$ is the minimal weight in $V_i$ (for $i=1,2$);
\item For each $\sigma \in \Sigma$, the set $\delta$ contains all the triples 
$\tuple{(q_1, v_1, q_2, v_2),\sigma,(q'_1,v'_1,q'_2,v'_2)}$ such that 
$v_i \in V_i$, $(q_i,\sigma,q'_i) \in \delta_i$, and $v'_i = \max\{v_i,\weight(q_i,\sigma,q'_i)\}$,
for $i=1,2$;
\item $\weight$ is defined by $\weight(\tuple{(q_1, v_1, q_2, v_2),\sigma,(q'_1,v'_1,q'_2,v'_2)}) = \min\{v'_1,v'_2\}$
for each $\tuple{(q_1, v_1, q_2, v_2),\sigma,(q'_1,v'_1,q'_2,v'_2)} \in \delta$.
\end{itemize}
If $A_1$ and $A_2$ are deterministic, then $A_{12}$ is deterministic.
The result for \dmax\/ follows.
\end{myProof}

\begin{theorem}
The (non)deterministic $\LimInf$-automata are closed under $\min$ with cost $O(n_1 \cdot n_2)$, and
the nondeterministic $\LimSup$-automata with cost $O(n_1\cdot n_2 \cdot (m_1 + m_2))$.
\end{theorem}

\begin{myProof}
Let $A_1=\tuple{Q_1,q_I^1,\Sigma,\delta_1,\weight_1}$ and $A_2=\tuple{Q_2,q_I^2,\Sigma,\delta_2,\weight_2}$
be two \nls. We construct a \nls\/ $A=\tuple{Q,q_I,\Sigma,\delta,\weight}$ such that
$L_A = \min\{L_{A_1},L_{A_2}\}$. Let $V_i = \{\weight_i(e) \mid e \in \delta_i\}$
be the set of weights that appear in $A_i$ (for $i=1,2$). Let $V_1 \cup V_2 = \{v_1,\dots,v_n\}$
and define
\begin{itemize}
\item $Q = \{q_I\} \cup Q_1 \times Q_2 \times \{1,2\} \times (V_1 \cup V_2)$ (where $q_I \not\in Q_1 \cup Q_2$ 
is a new state). Initially, a guess is made of the value $v$ of the input word.
Then, we check that both $A_1$ and $A_2$ visit a weight at least $v$ infinitely often.
In a state $\tuple{q_1,q_2,j,v}$ of $A$, the guess is stored in $v$ (and will never change
along a run) and the value of the index $j$ is toggled to $3-j$ as soon as $A_j$ 
does visit a weight at least $v$;
\item For each $\sigma \in \Sigma$, the set $\delta$ contains all the triples 
	\begin{itemize}
	\item $(q_I,\sigma,\tuple{q_1,q_2,1,v})$ such that 
		$v\in V_1 \cup V_2$ and for all $i \in \{1,2\}$, 
		we have $(q_I^{i}, \sigma, q_i) \in \delta_i$.
	\item $(\tuple{q_1,q_2,j,v}, \sigma, \tuple{q'_1,q'_2,j',v'})$ such that
		$v'=v$, $(q_i,\sigma,q'_i) \in \delta_i$ ($i=1,2$), and $j' = 3-j$ if $\weight_j(q_j,\sigma,q'_j) \geq v$,
        	and $j' = j$ otherwise. 
	\end{itemize}
\item $\weight$ is defined by $\weight(q_I,\sigma,\tuple{q_1,q_2,1,v}) = 0$ and $\weight(\tuple{q_1,q_2,j,v}, \sigma, \tuple{q'_1,q'_2,j',v'})$
is $v$ if $j \neq j'$ and $v_{\min}$ otherwise, where $v_{\min}$ is the minimal weight in $V_1 \cup V_2$.
\end{itemize}

For \dli, the construction is similar to the one presented in 
the proof of \theoremname~\ref{theo:max-closure} for \dls, where $\max$ is 
replaced by $\min$. The result for \nli\/ follows from the fact that 
\nli\/ is reducible to \dli.
\end{myProof}

\begin{theorem}
The deterministic $\LimSup$-automata are closed under $\min$ with cost $O(n_1 \cdot n_2 \cdot 2^{m_1 + m_2})$.
\end{theorem}

\begin{myProof}
Let $A_1=\tuple{Q_1,q_I^1,\Sigma,\delta_1,\weight_1}$ and $A_2=\tuple{Q_2,q_I^2,\Sigma,\delta_2,\weight_2}$
be two \dls. We construct a \dls\/ $A=\tuple{Q,q_I,\Sigma,\delta,\weight}$ such that
$L_A = \min\{L_{A_1},L_{A_2}\}$. Let $V_i = \{\weight_i(e) \mid e \in \delta_i\}$
be the set of weights that appear in $A_i$ (for $i=1,2$). For each weight
$v \in V_1 \cup V_2 = \{v_1,\dots,v_n\}$, we construct a \dbw\/ $A^v_{12}$ with accepting edges. 
The automaton $A^v_{12}$ consists of a copy of $A_1$ and a copy of $A_2$.
We switch from one copy to the other whenever an edge with weight at least $v$ 
is crossed. All such switching edges are accepting in $A^v_{12}$. The automaton $A$ then
consists of the synchronized product of these \dbw, where the weight of a joint
edge is the largest weight $v$ for which the underlying edge in $A^v_{12}$ is accepting.
Formally, let
\begin{itemize}
\item $Q = Q_1 \times Q_2 \times \{1,2\}^m$ where $m = \abs{V_1 \cup V_2}$;
\item $q_I = (q_I^1, q_I^2, b_1, \dots, b_m)$ where $b_i=1$ for all $1 \leq i \leq m$;
\item $\delta$ contains all the triples 
$(\tuple{q_1,q_2,b_1,\dots,b_m},\sigma, \tuple{q'_1,q'_2,b'_1,\dots,b'_m})$ such that
$\sigma \in \Sigma$ and 
	\begin{itemize}
	\item $(q_i,\sigma,q'_i) \in \delta_i$ for $i=1,2$;
	\item for all $1\leq j \leq m$, we have $b'_{j} = 3-b_j$ if 
$\weight_{b_j}(q_{b_j},\sigma,q'_{b_j}) \geq v_j$, and $b'_j = b_j$ otherwise.
	\end{itemize}
\item $\weight$ assigns to each transition 
$(\tuple{q_1,q_2,b_1,\dots,b_m},\sigma, \tuple{q'_1,q'_2,b'_1,\dots,b'_m}) \in \delta$ 
the weight $v=\max(\{v_{\min}\}\cup \{v_j \mid b_j\neq b'_j\})$
where $v_{\min}$ is the minimal weight in $V_1 \cup V_2$.
\end{itemize}

\end{myProof}

On the negative side, the (deterministic or not) limit-average and discounted-sum 
automata are not closed under $\min$. The following lemma establishes the
result for limit average.

\begin{lemma}\label{lem:limavg-min-comp}
Consider the alphabet $\Sigma=\set{a,b}$, and consider the languages
$L_a$ and $L_b$ that assigns the long-run average number 
of $a$'s and $b$'s, respectively.
Then the following assertions hold.
\begin{enumerate}
\item There is no \nla\/ for the language $L_m=\min\set{L_a,L_b}$.
\item There is no \nla\/ for the language $L^*= 1 -\max\set{L_a,L_b}$.
\end{enumerate}
\end{lemma}

\begin{myProof}
To obtain a contradiction, assume that there exists a \nla $A$ (for either $L_m$ or $L^*$). 
We first claim that there must be either an $a$-cycle or a 
$b$-cycle $C$ that is reachable in $A$ such that the sum of the weights in $C$ is positive.
Otherwise, if for all $a$-cycles and $b$-cycles we have that 
the sum of the weights is zero or negative, then we fool the automaton
as follows. Let $\beta$ be the maximum of the absolute values of the 
weights in $A$, and let $\alpha=\lceil \beta \rceil$. 
Then consider the word 
$w=(a^{5\cdot \alpha \cdot \abs{Q}} \cdot b^{5\cdot\alpha \cdot \abs{Q}})^\omega$.
For a run $r$ of $A$ over $w$, the long-run average of the
weights is bounded as follows:
\[
\frac{4 \cdot \beta \cdot \abs{Q}}{10\cdot \alpha\cdot \abs{Q}} \leq \frac{2}{5}.
\]
The above bound is as follows: in the run over $a^{5 \cdot\alpha \cdot \abs{Q}}$, there 
can be a prefix of size at most $\abs{Q}$ with sum of weights at most 
$\abs{Q}\cdot \beta$, and then there would be $a$-cycles, and then a trailing
prefix of size at most $\abs{Q}$ with sum of weights at most $\abs{Q}\cdot \beta$.
Similar argument holds for the segment of $b^{5 \cdot \alpha \cdot \abs{Q}}$.
Hence $L_{A}(w)\leq \frac{2}{5}$, however, $L_m(w)=L^*(w)=\frac{1}{2}$,
i.e., we have a contradiction.
W.l.o.g., we assume that there is an $a$-cycle $C$ such that sum of weights of $C$ 
is positive.
Then we present the following word $w$: a finite word $w_C$ to reach 
the cycle $C$, followed by $a^\omega$; the answer of the automaton is positive, 
{\it i.e.}, $L_{A}(w)>0$,  while $L_m(w)=L^*(w)=0$.
Hence the result follows.
\end{myProof}

\begin{theorem}\label{theo:dla-nla-not-closed-under-min}
The (non)deterministic $\LimAvg$-automata are not closed under $\min$.
\end{theorem}

\begin{myProof}
The result follows from \lemmaname~\ref{lem:limavg-min-comp} and the fact that 
there exists \dla\/ for the languages $L_a$ and $L_b$ of \lemmaname~\ref{lem:limavg-min-comp}.
\end{myProof}

Finally, we show that discounted-sum 
automata are not closed under $\min$.

\newcommand{\disc}{\mathit{discount}}

\begin{theorem}\label{theo:disc-min}
The (non)deterministic $\Disc$-automata are not closed under $\min$.
\end{theorem}

\begin{myProof}
Let $\lambda$ be a non-algebraic number in $]\frac{1}{2},1[$. 
We consider the quantitative languages 
$L_a^\lambda$ and $L_b^\lambda$ that assign the $\lambda$-discounted sum 
of $a$'s and $b$'s, respectively. 
Formally, given a (finite or infinite) word $w = w_0 w_1 \dots \in \Sigma^* \cup \Sigma^{\omega}$, let
$$r_a(w) = \sum_{i\mid w_i = a}^{\abs{w}} \lambda^i \quad \text{ and } \quad 
r_b(w) = \sum_{i\mid w_i = b}^{\abs{w}} \lambda^i $$
be the $\lambda$-discounted sum of the $a$'s (resp. $b$'s) of $w$. 
Then, $L_a^\lambda(w) = r_a(w)$ and $L_b^\lambda(w) = r_b(w)$. 
These languages are definable by \ddi. 
We show that the language $L_m =\min(L_a^\lambda, L_b^\lambda)$ is not definable by a \ndi.

Assume towards contradiction that there is a \ndi\/ $A$ for $L_m$.
By \lemmaname~5 and~6 in~\cite{CDH08}, there exists an infinite word $w^\prec$
such that $r_a(w^\prec) = r_b(w^\prec)$.

Since $r_a(w^\prec) + r_b(w^\prec) = \frac{1}{1-\lambda}$,
we have $L_m(w^\prec) = \frac{1}{2(1-\lambda)}$ and this is the maximal
value of a word in $L_m(\cdot)$. 

The maximal value in the automaton $A$ can be obtained for a lasso-word 
of the form $w_1.(w_2)^{\omega}$ (where $w_1,w_2$ are finite words 
and $w_2$ is nonempty), as pure memoryless strategies exist in games 
over finite graphs with the objective to maximize the discounted sum 
of payoffs. Since the language of $A$ is $L_m$, the value of $w_1.(w_2)^{\omega}$
is $\frac{1}{2(1-\lambda)}$, and thus $r_a(w_1.(w_2)^{\omega}) = r_b(w_1.(w_2)^{\omega})$ by
a similar argument as above. This last condition can be written as
$$p_a(\lambda) + \frac{\lambda^{n_1}\cdot q_a(\lambda)}{1-\lambda^{n_2}} = p_b(\lambda) + \frac{\lambda^{n_1}\cdot q_b(\lambda)}{1-\lambda^{n_2}}$$
for some polynomials $p_a, p_b, q_a, q_b$ and integers $n_1 \geq 0$ and $n_2 > 0$,
or more simply as
\begin{equation}
(1-\lambda^{n_2})\cdot p(\lambda) + \lambda^{n_1}\cdot q(\lambda) = 0 \label{eq:polynomial-lambda}
\end{equation}
for some polynomials $p$ of degree $n_1-1$ and $q$ of degree $n_2-1$,
all of whose coefficients are either $1$ or $-1$. Equation~\eqref{eq:polynomial-lambda}
is not identically zero as either $(i)$ $n_1 = 0$ and it reduces to $q(\lambda) = 0$
or $(ii)$ $n_1 > 0$ and then $p$ has degree at least $0$ so that the term of degree 
zero is not null in~\eqref{eq:polynomial-lambda}. 

Therefore, $\lambda$ must be algebraic, a contradiction.
\end{myProof}

\subsection{Closure under complement for infinite words}


Most of the weighted automata are not closed under complement.
The next result is a direct extension of the boolean case.

\begin{theorem}\label{theo:max-liminf-limsup-not-closed-under-complement}
The (non)deterministic $\Max$- and $\LimInf$-automata, and the deterministic 
$\LimSup$-automata are not closed under complement.
\end{theorem}

\begin{myProof}
The result follows from a similar result for the boolean version of these classes.
For \dmax\/ and \nmax, consider the language $L_1$ over $\Sigma=\{a,b\}$ 
such that $L_1(a^{\omega}) = 0$ and $L_1(w) = 1$ for all $w \neq a^{\omega}$.
For \dli\/ and \nli, consider the language $L_2$ over $\Sigma=\{a,b\}$ 
such that $L_2(\Sigma^*.a^{\omega}) = 1$ and $L(w) = 0$ for all words $w$ containing 
infinitely many $b$'s, and for \dls, consider $L_3$ the complement of $L_2$.
\end{myProof}

The next theorem is a positive result of closure under complementation
for \nls. It reduces to the complementation of nondeterministic B\"uchi automata.

\begin{theorem}\label{theo:nls-closed-under-complement}
The nondeterministic $\LimSup$-automata are closed under complement, with cost $O(m \cdot 2^{n \log n})$.
\end{theorem}

\begin{myProof}
Let $A=\tuple{Q,q_{0},\Sigma,\delta,\weight}$ be a \nls,
and let $V = \{\weight(e) \mid e \in \delta\}$
be the set of weights that appear in $A$.
For each $v \in V$, it is easy to construct a \nbw\/ $A_v$
whose (boolean) language is the set of words $w$
such that $L_A(w) \geq v$, by declaring to be accepting
the edges with weight at least $v$.
We then construct for each $v \in V$ a \nbw\/ $\bar{A}_v$ (with accepting edges) 
that accepts the (boolean) complement of the language accepted by $A_v$.
Finally, assuming that $V= \{v_1,\dots,v_n\}$ with $v_1 < v_2 < \dots < v_n$,
we construct the \nls\/ $B_i$ for $i=2,\dots,n$ where $B_i$ is obtained
from $\bar{A}_{v_i}$ by assigning weight $-v_{i-1}$ to each accepting edges,
and $-v_n$ to all the other edges. The complement of $L_A$ is then
$\max\{L_{B_2},\dots,L_{B_n}\}$ which is accepted by a \nls\/ by \theoremname~\ref{theo:max-closure}.
\end{myProof}

\begin{theorem}
The deterministic $\Disc$-automata are closed under complement, with cost $O(n)$.
\end{theorem}

\begin{myProof}[Sketch]
It suffices to replace each weight $v$ of a \ddi\/ by $1-\lambda-v$
(where $\lambda$ is the discount factor) to obtain the \ddi\/ for
the complement. 
\end{myProof}

\begin{theorem}
The deterministic $\LimAvg$-automata are not closed under complement.
\end{theorem}

\begin{figure}[t]
   \begin{center}
      \unitlength=.8mm
\def\fsize{\normalsize}

\begin{picture}(36,40)(0,0)

{\fsize

\node[Nmarks=i](x0)(18,18){}

\drawloop[ELside=l, ELdist=1, ELside=l,loopCW=y, loopdiam=7, loopangle=90](x0){$a,1$}
\drawloop[ELside=r, ELdist=1.5, ELside=r,loopCW=n, loopdiam=7, loopangle=-90](x0){$b,0$}



}
\end{picture}
   \end{center}
  \caption{Deterministic Limit-average Automaton.}
  \label{figure:aut1}
\end{figure}
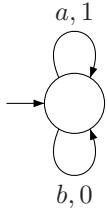 

\begin{myProof}
Consider the \dla\/ $A$ over alphabet $\Sigma= \{a,b\}$ 
(shown in \figurename~\ref{figure:aut1}) that consists of 
a single self-loop state with weight $1$ for $a$ and $0$ for $b$.
Notice that $L_A(w.a^\omega) = 1$ and $L_A(w.b^\omega) = 0$ for all $w \in \Sigma^*$.
To obtain a contradiction, assume that there exists a \dla\/ $B$ 
whose language is $L_B = 1-L_A$. For all finite words $w \in \Sigma^*$, 
let $L^{\Avg}_B(w)$ be the average weight of the unique (finite) run of $B$ over $w$.

Fix $0 < \epsilon < \frac{1}{2}$. For all finite words $w$, there exists
a number $n_w$ such that the average number of $a$'s in $w.b^{n_w}$ is at most $\epsilon$, 
and there exists a number $m_w$ such that $L^{\Avg}_B(w.a^{m_w}) \leq \epsilon$
(since $L_B(w.a^\omega) = 0$). Hence, we can construct a word 
$w = b^{n_1} a^{m_1} b^{n_2} a^{m_2} \dots$ such that 
$L_A(w) \leq \epsilon$ and $L_B(w) \leq \epsilon$.
Since $L_B = 1-L_A$, this implies that $1 \leq 2\epsilon$, a contradiction.
\end{myProof}

\begin{theorem}
The nondeterministic $\LimAvg$- and $\Disc$-automata are not closed under complement.
\end{theorem}

\begin{myProof}
The fact that \nla\/ are not closed under complementation is
as follows:
it follows from \lemmaname~\ref{lem:limavg-min-comp} that the 
language $L^*=1-\max\set{L_a,L_b}$ cannot be expressed as a 
\nla, however, the language $\max\set{L_a,L_b}$ can be expressed
as \nla\/ by \theoremname~\ref{theo:max-closure}.
That \ndi\/  are not closed under complement can be 
obtained as follows: given $0<\lambda<1$, consider the 
language $L_a^\lambda$ and $L_b^\lambda$ that assigns to words the
$\lambda$-discounted sum of $a$'s and $b$'s, respectively.
The language $L_a^\lambda$ and $L_b^\lambda$ can be expressed
as \ddi, and the max of them can be defined by \ndi.
Observe that $L_a^\lambda(w)+L_b^\lambda(w) = \frac{1}{1-\lambda}$ 
for all $w \in \Sigma^{\omega}$. Therefore, 
$\min\set{L_a^\lambda,L_b^\lambda}=\frac{1}{1-\lambda}-\max\set{L_a^\lambda,L_b^\lambda}$.
Since \ndi\/ is not closed under min (\theoremname~\ref{theo:disc-min}), we 
immediately obtain that \ndi\/ are not closed under complementation.
\end{myProof}

\subsection{Closure under sum for infinite words}

All weighted automata are closed under sum,
except \dla\/ and \nla.

\begin{theorem}\label{theo:max-closed-under-sum}
The (non)deterministic $\Max$-automata are closed under sum, with cost $O(n_1\cdot m_1 \cdot n_2 \cdot m_2)$.
\end{theorem}

\begin{myProof}[Sketch]
The construction in the proof of \theoremname~\ref{theo:max-closed-under-min}
can be adapted as follows: define the weight $\weight(\tuple{(q_1, v_1, q_2, v_2),\sigma,(q'_1,v'_1,q'_2,v'_2)})$
as $v'_1 + v'_2$ for each $\tuple{(q_1, v_1, q_2, v_2),\sigma,(q'_1,v'_1,q'_2,v'_2)} \in \delta$.
\end{myProof}

\begin{theorem}\label{theo:nls-closed-under-sum}
The nondeterministic $\LimSup$-automata are closed under sum, with cost $O(n_1\cdot m_1 \cdot n_2 \cdot m_2)$.
\end{theorem}

\begin{myProof}[Sketch]
Given two \nls\/ $A_1$ and $A_2$, we construct a \nls\/ $A$
for the sum of their languages as follows. Initially, we make a guess of
a pair $(v_1,v_2)$ of weights ($v_i$ in $A_i$, for $i=1,2$) and we branch
to a copy of the synchronized product of $A_1$ and $A_2$. We attach a bit $b$
whose range is $\{1,2\}$ to each state to remember that we expect $A_b$ to 
visit the guessed weight $v_b$. Whenever this occurs, the bit $b$ is set to $3-b$,
and the weight of the transition is $v_1 + v_2$. All other transitions ({\it i.e.}
when $b$ is unchanged) have weight $\min\{v_1 + v_2 \mid v_1 \in V_1 \land v_2 \in V_2\}$.
\end{myProof}

\begin{theorem}
The deterministic $\LimSup$-automata are closed under sum, with cost $O(n_1\cdot n_2 \cdot 2^{m_1 \cdot m_2})$.
\end{theorem}

\begin{myProof}
Let $A_1=\tuple{Q_1,q_I^1,\Sigma,\delta_1,\weight_1}$ and $A_2=\tuple{Q_2,q_I^2,\Sigma,\delta_2,\weight_2}$
be two \dls. We construct a \dls\/ $A=\tuple{Q,q_I,\Sigma,\delta,\weight}$ such that
$L_A = L_{A_1} + L_{A_2}$. Let $V_i = \{\weight_i(e) \mid e \in \delta_i\}$
be the set of weights that appear in $A_i$ (for $i=1,2$). 
The automaton $A$ implements the synchronized product of $A_1$ and $A_2$,
and keeps one bit $b(v_1,v_2)$ for each pair $(v_1,v_2)$ of weights $v_1 \in V_1$ and $v_2 \in V_2$.
For $i=1,2$, if $b(v_1,v_2)=i$, then $A_i$ is expected to cross a transition with weight
$v_i$. Whenever this occurs, the bit is set to $3-i$.
The weight of a transition in $A$ is the largest value of $v_1 + v_2$
such that the corresponding bit $b(v_1,v_2)$ has changed in the transition.
Formally, we define:
\begin{itemize}
\item $Q = Q_1 \times Q_2 \times [V_1 \times V_2 \to \{1,2\}]$;
\item $q_I = \tuple{q_I^1,q_I^2,b_I}$ where $b_I(v_1,v_2) = 1$ for all $(v_1,v_2) \in V_1 \times V_2$;
\item For each $\sigma \in \Sigma$, the set $\delta$ contains all the triples 
 $(\tuple{q_1,q_2,b}, \sigma, \tuple{q'_1,q'_2,b'})$ such that
	$(q_i,\sigma,q'_i) \in \delta_i$ ($i=1,2$), and for all $(v_1,v_2) \in V_1 \times V_2$, 
	we have $b'(v_1,v_2) = 3-b(v_1,v_2)$ 
	if $\weight_i(\tuple{q_i,\sigma,q'_i}) = v_i$ for $i=b(v_1,v_2)$,
	and otherwise $b'(v_1,v_2) = b(v_1,v_2)$.
\item $\weight$ is defined by $\weight(\tuple{q_1,q_2,b}, \sigma, \tuple{q'_1,q'_2,b'}) = 
\max(\{v_{\min} \cup \{v_1 + v_2 \mid b'(v_1,v_2) \neq b(v_1,v_2) \})$
where $v_{\min}$ is the minimal weight in $V_1 + V_2 = \{v_1 + v_2 \mid v_1 \in V_1 \land v_2 \in V_2\}$.
\end{itemize}
\end{myProof}

\begin{theorem}
The (non)deterministic $\LimInf$-automata are closed under sum with cost $O(n_1\cdot n_2 \cdot 2^{m_1 \cdot m_2})$.
\end{theorem}

\begin{myProof}
Let $A_1=\tuple{Q_1,q_I^1,\Sigma,\delta_1,\weight_1}$ and $A_2=\tuple{Q_2,q_I^2,\Sigma,\delta_2,\weight_2}$
be two \dli. We construct a \dli\/ $A=\tuple{Q,q_I,\Sigma,\delta,\weight}$ such that
$L_A = L_{A_1} + L_{A_2}$. Let $V_i = \{\weight_i(e) \mid e \in \delta_i\}$
be the set of weights that appear in $A_i$ (for $i=1,2$). 
The automaton $A$ implements the synchronized product of $A_1$ and $A_2$,
and keeps one bit $b(v_1,v_2)$ for each pair $(v_1,v_2)$ of weights $v_1 \in V_1$ and $v_2 \in V_2$.
If a transition in $A_i$ for some $i \in \{1,2\}$ has weight less than $v_i$,
then the bit $b(v_1,v_2)$ is set to $\bot$, otherwise is set to $\top$.
The weight of a transition in $A$ is the largest value of $v_1 + v_2$
such that the corresponding bit $b(v_1,v_2)$ is $\top$.
Formally, we define:
\begin{itemize}
\item $Q = Q_1 \times Q_2 \times [V_1 \times V_2 \to \{\top,\bot\}]$;
\item $q_I = \tuple{q_I^1,q_I^2,b_I}$ where $b_I(v_1,v_2) = \bot$ for all $(v_1,v_2) \in V_1 \times V_2$;
\item For each $\sigma \in \Sigma$, the set $\delta$ contains all the triples 
 $(\tuple{q_1,q_2,b}, \sigma, \tuple{q'_1,q'_2,b'})$ such that
	$(q_i,\sigma,q'_i) \in \delta_i$ ($i=1,2$), and for all $(v_1,v_2) \in V_1 \times V_2$, 
	we have $b'(v_1,v_2) = \top$ 
	if $\weight_i(\tuple{q_i,\sigma,q'_i}) \geq v_i$ for $i=1,2$,
	and otherwise $b'(v_1,v_2) = \bot$.
\item $\weight$ is defined by $\weight(\tuple{q_1,q_2,b}, \sigma, \tuple{q'_1,q'_2,b'}) = 
\max(\{v_{\min} \cup \{v_1 + v_2 \mid b'(v_1,v_2) = \top \})$
where $v_{\min}$ is the minimal weight in $V_1 + V_2 = \{v_1 + v_2 \mid v_1 \in V_1 \land v_2 \in V_2\}$.
\end{itemize}
The result for \nli\/ follows from the fact that \nli\/ is reducible to \dli.
\end{myProof}

\begin{theorem}\label{theo:ddi-ndi-closed-under-sum}
The (non)deterministic $\Disc$-automata are closed under sum, with cost $O(n_1 \cdot n_2)$.
\end{theorem}

\begin{myProof}[Sketch]
It is easy to see that the synchronized product of two \ndi\/ (resp. \ddi)
defines the sum of their languages, if the weight of a joint transition
is defined as the sum of the weights of the corresponding transitions in the two \ndi\/ (resp. \ddi).
\end{myProof}

\begin{theorem}
The (non)deterministic $\LimAvg$-automata are not closed under sum.
\end{theorem}

\begin{myProof}
Consider the alphabet $\Sigma=\set{a,b}$, and consider the \dla-definable languages 
$L_a$ and $L_b$ that assigns to each word $w$ the long-run average number of $a$'s and $b$'s in $w$
respectively. 
Let $L_{+}=L_a + L_b$. Assume that $L_{+}$ is defined by a \nla\/ $A$ with set 
of states $Q$ (we assume w.l.o.g that every state in $Q$ is reachable).

First, we claim that from every state $q \in Q$, there is a run of $A$ over $a^{\abs{Q}}$
that visit a cycle $C^*$ with average weight $1$. To see this, notice that from every state $q \in Q$,
there is an infinite run $\rho$ of $A$ over $a^\omega$ whose value is $1$ 
(since $L_{+}(w_q\cdot a^\omega) =1$ for all finite words $w_q$). 
Consider the following decomposition of $\rho$.
Starting with an empty stack, we push the states of $\rho$ onto the stack
as soon as all the states on the stack are different. If the next state is already
on the stack, we pop all the states down to the repeated state thus removing a simple
cycle of $\rho$. Let $C_1$, $C_2, \dots$ be the cycles that are successively 
removed. Observe that the height of the stack is always at most $\abs{Q}$.
Let $\beta$ be the largest average weight of the cycles $C_i$, $i\geq 1$,
and let $\alpha_{\max}$ be the largest weight in $A$.
Assume towards contradiction that $\beta < 1$. Then, for all $n > 0$, 
the value of the prefix of length $n$ of $\rho$ is at most:
$$ \frac{\alpha_{\max} \cdot \abs{Q} + \beta \cdot \sum_{i=1}^{k_n} \abs{C_i}}{n} $$
where $k_n$ is the number of cycles that have been removed from the stack 
when reading the first $n$ symbols of $\rho$. Hence, the value of $\rho$ is at most $\beta < 1$,
which is a contradiction. Therefore, the average weight of some cycle $C^* = C_i$ 
is exactly\footnote{It cannot be greater than $1$ since $L_{+}(w\cdot a^\omega)=1$ for all finite 
words $w$.} $1$ (there are finitely many different cycles as they are simple cycles).
Since the height of the stack is at most $\abs{Q}$, the cycle $C^*$ is reachable
in at most $\abs{Q}$ steps.

Second, it can be shown analogously that from every state $q \in Q$, there is a 
run over $b^{\abs{Q}}$ that visit a cycle $C^*$ with average weight $1$.

Third, for arbitrarily small $\epsilon >0$, consider the word $w$ and the run $\rho$ of $A$ over $w$ generated inductively 
by the following procedure: $w_0$ is the empty word and $\rho_0$ is the initial state of $A$
We generate $w_{i+1}$ and $\rho_{i+1}$ from $w_i$ and $\rho_i$ as follows:
\begin{compressEnum}
\itCompress generate a long enough sequence $w_{i+1}'$ of $a$'s after $w_i$ such that the 
average number of $b$'s in $w_i \cdot w_{i+1}'$ falls below $\epsilon$
and we can continue $\rho_i$ and reach within at most $\abs{Q}$ steps (and then repeat $k$ times) 
a cycle $C$ of average weight $1$ and such that the average weight of this run
prolonged by $\abs{Q}$ arbitrary transitions is at least $1-\epsilon$, {\it  i.e.}
$$ \frac{\gamma(\rho_i)+k\cdot\abs{C} + 2 \alpha_{\min} \cdot \abs{Q} }{\abs{\rho_i} + k\cdot \abs{C} + 2 \cdot \abs{Q}}
\geq 1-\epsilon$$
where $\alpha_{\min}$ is the least weight in $A$. This is possible since $k$ can be chosen
arbitrarily large. Let $\rho'_i$ be the prolongation of $\rho_i$ over $w_{i+1}'$;
\itCompress then generate a long enough sequence $w_{i+1}''$ of $b$'s such that the average number of 
$a$'s in $w_{i} \cdot w_{i+1}' \cdot w_{i+1}''$ falls below $\epsilon$
and as above, we can construct a continuation $\rho''_i$ of $\rho'_i$ whose
average weight is at least $1-\epsilon$ (even if prolonged by $\abs{Q}$ arbitrary transitions);
\itCompress the word $w_{i+1}=w_i \cdot w_{i+1}' \cdot w_{i+1}''$ and the run 
$\rho_{i+1}$ is $\rho''_i$.
\end{compressEnum}
The word $w$ and the run $\rho$ are the limit of these sequences.
We have $L_{a}(w) = L_{b}(w) = 0$ and thus $L_{+}(w) = 0$, while
the value of $\rho$ is at least $1-\epsilon$, a contradiction.
\end{myProof}

\paragraph{{\bf Acknowledgment.}} We thank Wolfgang Thomas for pointing out the isolated cut-point problem.

\bibliography{biblio}

\newcommand{\etalchar}[1]{$^{#1}$}
\begin{thebibliography}{CdAHS03}

\bibitem[CCH{\etalchar{+}}05]{CCHK+05}
A.~Chakrabarti, K.~Chatterjee, T.~A. Henzinger, O.~Kupferman, and R.~Majumdar.
\newblock Verifying quantitative properties using bound functions.
\newblock In {\em CHARME}, LNCS 3725, pages 50--64. Springer, 2005.

\bibitem[CdAHS03]{CAHS03}
A.~Chakrabarti, L.~de~Alfaro, T.~A. Henzinger, and M.~Stoelinga.
\newblock Resource interfaces.
\newblock In {\em EMSOFT}, LNCS 2855, pages 117--133. Springer, 2003.

\bibitem[CDH08]{CDH08}
K.~Chatterjee, L.~Doyen, and T.~A. Henzinger.
\newblock Quantitative languages.
\newblock In {\em CSL}, LNCS 5213, pages 385--400. Springer, 2008.

\bibitem[CGH{\etalchar{+}}08]{CGHIKPS08}
K.~Chatterjee, A.~Ghosal, T.~A. Henzinger, D.~Iercan, C.~Kirsch, C.~Pinello,
  and A.~Sangiovanni-Vincentelli.
\newblock Logical reliability of interacting real-time tasks.
\newblock In {\em DATE}, pages 909--914. ACM, 2008.

\bibitem[Cha07]{Cha-TCS}
K.~Chatterjee.
\newblock {\em Stochastic $\omega$-Regular Games}.
\newblock PhD thesis, University of California, Berkeley, 2007.

\bibitem[CK94]{CulikK94}
Karel {Culik II} and Juhani Karhum{\"a}ki.
\newblock Finite automata computing real functions.
\newblock {\em SIAM J. Comput.}, 23(4):789--814, 1994.

\bibitem[CM00]{CortesM00}
Corinna Cortes and Mehryar Mohri.
\newblock Context-free recognition with weighted automata.
\newblock {\em Grammars}, 3(2/3):133--150, 2000.

\bibitem[Con92]{Condon92}
Anne Condon.
\newblock The complexity of stochastic games.
\newblock {\em Inf. Comput.}, 96(2):203--224, 1992.

\bibitem[dAHM03]{AHM03}
L.~de~Alfaro, T.~A. Henzinger, and R.~Majumdar.
\newblock Discounting the future in systems theory.
\newblock In {\em ICALP}, LNCS 2719, pages 1022--1037. Springer, 2003.

\bibitem[DG07]{DrosteGastin07}
M.~Droste and P.~Gastin.
\newblock Weighted automata and weighted logics.
\newblock {\em Th. C. Sci.}, 380(1-2):69--86, 2007.

\bibitem[DK03]{DrosteK03}
Manfred Droste and Dietrich Kuske.
\newblock Skew and infinitary formal power series.
\newblock In {\em ICALP}, LNCS 2719, pages 426--438. Springer, 2003.

\bibitem[DKR08]{DrosteKR08}
Manfred Droste, Werner Kuich, and George Rahonis.
\newblock Multi-valued {MSO} logics over words and trees.
\newblock {\em Fundamenta Informaticae}, 84(3-4):305--327, 2008.

\bibitem[DR07]{DrosteR07}
Manfred Droste and George Rahonis.
\newblock Weighted automata and weighted logics with discounting.
\newblock In {\em CIAA}, LNCS 4783, pages 73--84. Springer, 2007.

\bibitem[{\'E}K04]{EsikK04}
Zolt{\'a}n {\'E}sik and Werner Kuich.
\newblock An algebraic generalization of omega-regular languages.
\newblock In {\em MFCS}, LNCS 3153, pages 648--659. Springer, 2004.

\bibitem[EM79]{EM79}
A.~Ehrenfeucht and J.~Mycielski.
\newblock Positional strategies for mean payoff games.
\newblock {\em Int. Journal of Game Theory}, 8(2):109--113, 1979.

\bibitem[GC03]{GurfinkelC03}
Arie Gurfinkel and Marsha Chechik.
\newblock Multi-valued model checking via classical model checking.
\newblock In {\em CONCUR}, LNCS 2761, pages 263--277. Springer, 2003.

\bibitem[Kar78]{Karp78}
R.~M. Karp.
\newblock A characterization of the minimum cycle mean in a digraph.
\newblock {\em Discrete Mathematics}, 23(3):309--311, 1978.

\bibitem[KL07]{LatticeAutomata07}
O.~Kupferman and Y.~Lustig.
\newblock Lattice automata.
\newblock In {\em VMCAI}, LNCS 4349, pages 199--213. Springer, 2007.

\bibitem[KS86]{KuichS86}
Werner Kuich and Arto Salomaa.
\newblock {\em Semirings, Automata, Languages}, volume~5 of {\em EATCS
  Monographs in Theoretical Computer Science.}
\newblock Springer, 1986.

\bibitem[Rab63]{Rabin63}
Michael~O. Rabin.
\newblock Probabilistic automata.
\newblock {\em Information and Control}, 6(3):230--245, 1963.

\bibitem[Sch61]{Wautomata}
M.~P. Sch\"{u}tzenberger.
\newblock On the definition of a family of automata.
\newblock {\em Information and control}, 4(2-3):245--270, 1961.

\bibitem[Sha53]{Sha53}
L.~S. Shapley.
\newblock Stochastic games.
\newblock In {\em Proc. of the National Acadamy of Science USA}, volume~39,
  pages 1095--1100, 1953.

\bibitem[VW86]{VardiW86}
Moshe~Y. Vardi and Pierre Wolper.
\newblock An automata-theoretic approach to automatic program verification.
\newblock In {\em LICS}, pages 332--344. IEEE, 1986.

\bibitem[ZP96]{ZwickP96}
Uri Zwick and Mike Paterson.
\newblock The complexity of mean payoff games on graphs.
\newblock {\em Theor. Comput. Sci.}, 158(1{\&}2):343--359, 1996.

\end{thebibliography}
\bibliographystyle{alpha}
\end{document}